\documentclass[preprints,article,accept,moreauthors,pdftex]{Definitions/mdpi}

\firstpage{1} 
\pubvolume{1}
\issuenum{1}
\articlenumber{0}
\pubyear{2025}
\copyrightyear{2025}
\datereceived{} 
\dateaccepted{} 
\datepublished{} 
\hreflink{https://doi.org/} 

\Title{Nonextensive aspects of gluon distribution and the implications to QCD phenomenology}

\TitleCitation{}

\Author{Lucas Moriggi $^{1,\dagger,\ddagger}$\orcidA{} and Magno Machado $^{2,\ddagger}$\orcidB{}}

\AuthorNames{Lucas Moriggi and Magno Machado}

\AuthorCitation{Moriggi, L.;  Machado, M.}

\address{%
$^{1}$ \quad Universidade Estadual do Centro-Oeste (UNICENTRO), Guarapuava, PR, Brazil; lucasmoriggi@unicentro.br\\
$^{2}$ \quad Instituto de Física, Universidade Federal do Rio Grande do Sul, Porto Alegre, RS, Brazil; magnus@if.ufrgs.br}

\abstract{This study presents new insights into gluon transverse momentum distributions through nonextensive statistical mechanics, addressing their implications for QCD phenomenology. The saturation physics and scaling laws present in high energy collision data are investigated as a consequence of gluon distribution  modification at high density regime. The analysis explores how these modifications  influence observables across different collision systems, such as proton-proton, proton-nucleus, and relativistic heavy-ion collisions. Both high and low $p_T$ regions are successfully described in hadron production.}

\keyword{$k_T$-factorization approach; parton saturation phenomenon; geometric scaling in hadron-hadron collisions; high-energy collisions; transversal momentum distribution; nonextensive statistical mechanics } 

\begin{document}

\section{Introduction}

At high energies, the proton can be viewed as a system with a high density of gluons, making the gluon distribution a fundamental aspect in the description of both hard and soft processes at the Large Hadron Collider (LHC) energies. Although the traditional collinear factorization framework, based on perturbative resummation of large logarithms of the scale $Q^2$, provides good results for processes characterized by a large hard scale, this approach has limitations in several contexts. Initially, it was expected that the hard-scale dynamics of Quantum Chromodynamics (QCD) at high energies would simplify, allowing collinear factorization to generate highly accurate predictions. However, novel effects arising from cold and hot nuclear matter in TeV-scale collisions have challenged this assumption.

High-multiplicity events in proton-proton ($pp$) collisions exhibit characteristics that are difficult to reconcile with collinear factorization, particularly at moderate or small transverse momentum $p_T$. Similar challenges occur in hadron production in proton-nucleus ($pA$) collisions, where cold nuclear matter effects become significant and there is a need for additional parameterization of nuclear parton distribution functions (PDFs) to fit experimental data. The most complex scenario arises in heavy-ion collisions ($AA$), where collective effects and quark-gluon plasma thermalization lead to notable deviations from the collinear model. Furthermore, small-$x$ processes, such as diffractive production, require alternative factorization mechanisms, even when a semi-hard momentum scale is involved.

Given these limitations, a more appropriate description of the proton is expected to come from $k_T$-factorization \cite{Catani:1990xk,Lipatov:2013yra,Lipatov:2022doa,Boer:2016bfj,Cisek:2014ala} where the parton distribution depends on the transverse momentum of its constituents $\phi(x, k_t, Q^2)$. Various models have been proposed for these distributions based on different physical assumptions \cite{Abdulov:2021ivr,Moriggi:2020zbv,Bacchetta:2020vty,Lipatov:2022doa,Angeles-Martinez:2015sea,Abdulov:2021ivr,Luszczak:2022fkf,GBWnovo,Kutak:2012rf}.

This work investigates thermal aspects of gluon distributions, especially at high energies, and their implications for calculating observables in different systems, including electron-proton ($ep$), 
 electron-nucleus ($eA$), $pp$, and $AA$ collisions. In previous studies \cite{Moriggi:2024tbr}, it was demonstrated that the transverse-momentum-dependent gluon distribution, parametrized as a power law \cite{Moriggi:2020zbv}, can be interpreted using non-extensive statistical mechanics, which was used to extract information about high-multiplicity processes at the LHC. It was shown that both the parameter interpreted as temperature and the power-law index associated with Tsallis entropy $q$ can be parametrized in terms of partonic variables $x$ and $Q^2$, resulting in cross-sections that exhibit scaling behavior. These scaling laws can then be employed to predict gluon distributions and other observables.

The importance of such investigations lies in the fact that parton distributions are essential for determining any observable in hadronic processes. Precise knowledge of these quantities reduces uncertainties in Standard Model predictions and allows for a deeper understanding of QCD dynamics in different regimes. Moreover, controlling the nuclear modifications of parton distributions is crucial for heavy-ion physics and the study of quark-gluon plasma.

Experimental data for different observables, in a wide range of partonic variables ($x, k_T$), support the scaling hypothesis, where data from different collider energies $\sqrt{s}$ and hard scales $Q^2$ collapse into a single scaling line related to the variable $\tau = Q^2/Q_s^2(x)$ \cite{Moriggi:2020zbv,McLerran:2014apa,Stasto,Praszalowicz:2011tc,Praszalowicz:2013fsa,Praszalowicz:2015dta}. The onset of the gluon saturation phenomenon is driven by a dynamically generated transverse momentum scale, the saturation scale $Q_s$. The growth of cross-sections as a function of interaction area and multiplicity can also follow this scaling by varying the scale $Q_s(x)$ with multiplicity \cite{Osada:2019oor,Osada:2020zui,Moriggi:2024tbr}. Furthermore, hadron production in the $p_T$ spectra is characterized by a power law, partially attributed to the behavior of parton distributions.

In Ref. \cite{Moriggi:2020zbv}, we propose a model that parameterizes the gluon distributions, incorporating scaling and a power law parameter characterizing the high-momentum spectrum. This parameterization has been used to describe diffractive processes \cite{Peccini:2020jkj,Peccini:2021rbt,Cisek:2022yjj,Luszczak:2022fkf,SampaiodosSantos:2021tfh,Santos:2023yep}, $p_T$ spectra in $pp$ collisions \cite{Moriggi:2020zbv}, and heavy-ion physics \cite{Moriggi:2020qla,Moriggi:2023ahi}. In \cite{Moriggi:2024tbr}, a connection with non-extensive statistical mechanics was established, showing that experimental data on multiplicity variation can be interpreted by analyzing partonic entropy and its dependence on the transverse overlap area of protons.

Although the thermal properties of transverse momentum hadronic spectra have been explored for long time in electron-positron ($e^+ e^-$), $pp$, $pA$, $AA$ collisions \cite{Hagedorn,Wong:2015mba,Bhattacharyya:2017cdk,Biro:2017arf, Biro:2020kve,Akhil:2023xpb,Li:2020lww,Sharma:2018jqf,Khuntia:2017ite,Bhattacharyya:2016lrk,Parvan:2016rln,Tripathy:2016hlg}, our approach investigates the thermal like appearance of the spectra as a consequence of transverse momentum distribution (TMD) and $k_T$- factorization which attribute an universal behavior that can be used to generate predictions to different observables and improve our understanding of soft aspects of QCD. 

One of key ideas of using the statistical mechanics description is that the maximum entropy method can be used to get some information about the hadron structure. For instance, in Refs. \cite{Han:2018wsw,Chen:2024dhz} the authors determine the pion valence quark content at low resolution $Q^2\sim 0.1$ GeV$^2$ using the maximum entropy plus DGLAP evolution with non-linear correction \cite{GLR,MQ}. Moreover, entanglement entropy has been used as a tool to analyze QCD at soft scales and color confinement which can be used to understand hadron production in deep inelastic scattering and small-$x$ parton cascade \cite{Kharzeev:2017qzs,Hentschinski:2021aux,Hentschinski:2023izh,Hentschinski:2024gaa}.

Although new phenomenological results are presented in this article, we also provide a brief review of the topic and suggest possible future research directions. In Section \ref{sec:UGD}, we discuss how traditional PDFs described by DGLAP evolution at large $Q^2$ can be parametrized in a power-law form, offering insights into the non-perturbative aspects of these distributions. In Section \ref{sec:mult}, we calculate the hadronic spectrum for the production of pions and kaons as a function of the multiplicity class, demonstrating that both can be placed on a scaling line. In Section \ref{sec:DISeA}, we present a possible extension for nuclear cases, considering Glauber's multiple independent interaction model, and provide results for the nuclear modification factor in $eA$ collisions in the small-$x$ region. Finally, in Section \ref{sec:heavyion}, we discuss heavy-ion collisions, showing that an additional modification is necessary to describe the LHC data.

\section{Gluon distribution and power-like parametrization}
\label{sec:UGD}
The thermal interpretation of parton distributions offers a unified framework for describing the infrared regime of QCD. Traditionally, we are compelled to separate the non perturbative part of the PDF from that evolved by perturbative dynamics. So different frameworks are used to study these regimes. However, we can make use of the thermal picture to extend the unintegrated gluon distribution to the soft part of the partonic spectra. The accuracy of this extension can only be measured by the posterior ability to make sense of the data when using it. So, for now, we present it as a possibility.

Collinear PDFs \cite{MMHT2014,Hou:2019efy,Bailey:2020ooq,NNPDF:2017mvq,HERAPDF} are usually obtained as a result of global analysis of high $Q^2$ data where distributions evolve from the initial scale $Q_0^2$ to high $Q^2$ via Dokshitzer – Gribov – Lipatov – Altarelli – Parisi (DGLAP) evolution equations. The gluon integrated PDF, $xG(x,Q^2)$, can be related to the transverse momentum distribution (TMD) or unintegrated gluon distribution (UGD) in small-x limit as 
\begin{equation}\label{eq:integrated}
    \frac{\partial}{\partial k_T^2} xG(x,k_T) = \phi (x,k_T).  
\end{equation}
This relationship can be used to obtain an unintegrated counterpart of the distribution in the small-$x$ limit. TMDs are more sensitive to partonic dynamics than their integrated counterparts. The UGDs obtained from collinear PDF have power-like tails that result from gluon splitting and have an energy-dependent behavior. Let us define the rapidity of the gluon as $Y=\log(1/x)$ and then at some rapidity $Y_0$ we expected that the gluon distribution follows a behavior given by its leading-order value $\phi_{LO} (x_0,k_T) \sim 1/k_T^2$. However, the gluon splitting process in different approaches like DGLAP \cite{D,GL,AP}, 
 Balitskii-Fadin-Kuraev-Lipatov (BFKL) \cite{BFKL1,BFKL2}, Balitsky-Kovchegov (BK) \cite{BK1,BK2} changes this power-like behavior in the form $\phi(x,k_T)\sim 1/k_T^{2+2\delta n}$. 
 
 Building upon the foundational framework of unintegrated gluon distributions, we now examine their thermal interpretation, emphasizing power-like behavior and its connection to nonextensive statistical mechanics. The UGDs parametrizations are more clear in color dipole picture where the scattering probability of the dipole-proton interaction in momentum space, $P(x,k_T)$, is related with UGD in the following way:
\begin{equation}
\label{eq:phi-1}
\phi(x,k_T)= \frac{3\sigma_0}{4\pi^2 \alpha_s} k_T^2 P(x,k_T).
\end{equation}
The constant factor $\sigma_0$ is related to the transverse area of the proton and this relationship establishes that the fundamental element in the description of proton structure is the scattering probability.
As we proposed in Ref. \cite{Moriggi:2020zbv} the amplitude $P(x,k_T)$ can be write in the power-like form
\begin{equation}\label{eq:MPM}
P^{MPM}(\delta n,x,k_T)=\frac{1+\delta n}{\pi Q^2_s(x)} \frac{1}{\left(1+k_T^2/Q_s^2(x)\right)^{2+\delta n}}.
\end{equation}

In \cite{Moriggi:2024tbr}  we argue that Eq. \eqref{eq:MPM} can be obtained by maximizing the Tsallis entropy, $S_q$: \cite{Tsallis:1987eu,Tsallis:1998ws},
\begin{equation}\label{eq:qentro}
    S_q=\int d^2k_T \frac{1-\left[ P(x,k_T) \right]^{q}}{q-1}. 
\end{equation}
and we interpret the Lagrange parameter as the inverse of the temperature $\beta^{-1}=T$, using the scaling hypothesis:
\begin{equation}
    \langle k_T^2 (x) \rangle_q \sim \beta^{-1} (x_s/x)^{\lambda} ,
\end{equation}
where one has a generalization of Einstein relation for anomalous diffusion \cite{PhysRevE.60.2398,PhysRevLett.75.366}. Thus we can relate the quantities $\delta n$  and the gluon saturation scale $Q_s(x)$ to nonextensive quantities $T$, $q$:
\begin{equation} \label{eq:nepars}
    q=\frac{3+\delta n}{2 + \delta n}, \qquad T=Q_s^2 (q-1).
\end{equation}
From deep inelastic scattering (DIS) and inclusive pion production in $pp$ collision data we determine the following scaling form parametrization for these quantities (form more details, see \cite{Moriggi:2020zbv}):
\begin{equation}
\label{eq:pars}
  \delta n (Q^2/Q_s(x)^2) = 0.075 \, \left( \frac{Q^2}{Q_s^2(x)} \right)^{0.188} , \qquad Q_{s}^2(x)= \left( \frac{x_s}{x}\right) ^{1/3}.
\end{equation}

\begin{figure*}[t]
\includegraphics[width=\linewidth]{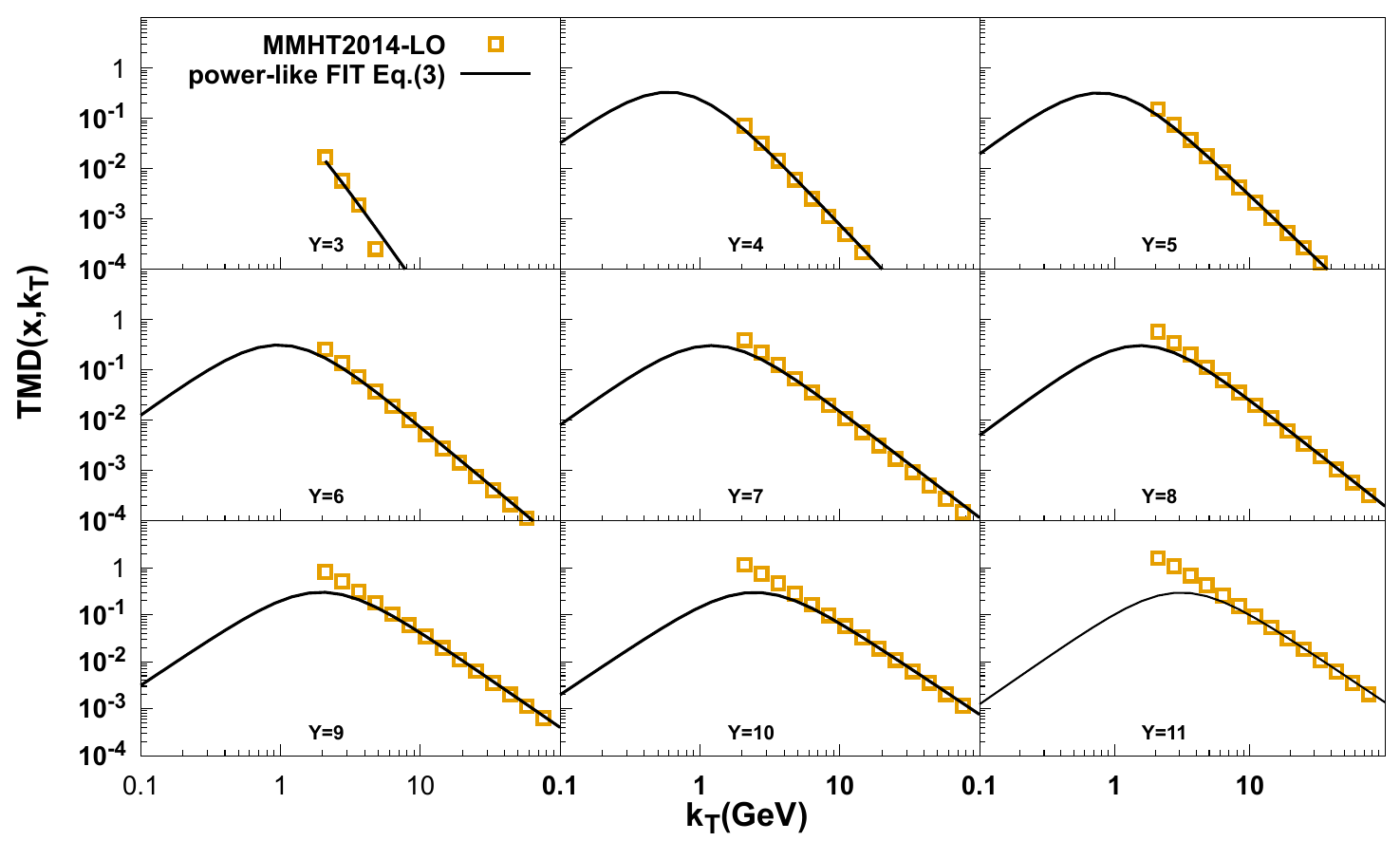}
\caption{The TMD $\phi(x,k_T)$ obtained from MMHT2014-LO collinear gluon PDF (open squares) by differentiation of the integrated distribution $xG(x,Q^2)$ from Eq. \eqref{eq:integrated} compared with its power-like fit using Eq.\eqref{eq:MPM} and \eqref{eq:phi-1}. }
\label{fig:MMHT2014} 
\end{figure*}

Now we focus on the discussion of collinear PDFs and their predicted behavior to these quantities. In order to investigate the non-extensive aspects of these distributions, we use MMHT-2014 gluon PDF in its leading order version \cite{MMHT2014}. Therefore, we can obtain the unintegrated version by its derivative \eqref{eq:integrated} for $k_T > Q_0=2$ GeV at different values of $Y$ as shown in Figure \ref{fig:MMHT2014} in open squares. At large $Y$ the distribution became less steep and close to its leading order value. 

We then apply the Eq. \eqref{eq:MPM} to calculate the UGD from Eq. \eqref{eq:phi-1} to fit the parameters $Q_s$ and $\delta n$ to the distribution at each selected value of $Y$. By doing this, we can extract the thermal parameters $T$ and $q$ from the MMHT2014 distribution considering the identification by Eq. \eqref{eq:nepars}. The fit results are presented in Fig. \ref{fig:MMHT2014} in full lines from high $k_T$ up to $k_T=0$ where the infrared limit was naturally regulated by the saturation of the distribution. We can notice that different models for the partonic distribution will be characterized by their own parameters $T$ and $q$ based on the theoretical assumptions that underlie each one. The resulting parameters at each value of $Y$ are presented in Table \ref{tab:pars} for $x$ in the range $2.3\times 10^{-6} - 5\times 10^{-2}$ and $k_T<100$ GeV. Within the range $x>10^{-4}$ where experimental data are available, there is no substantial distinction between the extrapolation and the true values.

\begin{table}[t]
\centering
\begin{tabular}{|c|c|c|}
\hline
\textbf{$Y$}  & \textbf{$T$} & \textbf{$\delta n$} \\ \hline
3.0   & 0.368  & 0.968  \\ \hline
4.0   & 0.516  & 0.466  \\ \hline
5.0   & 0.761  & 0.282  \\ \hline
6.0   & 1.083  & 0.157  \\ \hline
7.0   & 1.557  & 0.061  \\ \hline
8.0   & 2.523  & 0.059 \\ \hline
9.0   & 3.904  & 0.024 \\ \hline
10.0  & 6.021  & -0.006 \\ \hline
11.0  & 9.307  & -0.033 \\ \hline
12.0  & 14.39 & -0.0579 \\ \hline
13.0  & 22.41 & -0.0806 \\ \hline
\end{tabular}
\caption{Parameters (temperature $T$ and index $\delta n$) associated to the power-like form applied to the collinear MMHT2014-LO gluon distribution given by Eq. \eqref{eq:MPM} and \eqref{eq:phi-1} at different values of gluon rapidity $Y$.}
\label{tab:pars}
\end{table}

This procedure can give us a different interpretation of the collinear-like distribution in the infrared regime. First, we observe that the peak (maximum) moves towards high $k_T$ at small-$x$ where a large number of soft gluons dominate the proton wavefunction, indicating a larger temperature regime. At the same time $\delta n$ approaches the zero limit, which indicates the entropic index $q=3/2$. In large $k_T$ and $x< 10^{-5}$ where PDFs do not have data to be constrained, we observe that $\delta n$ slightly negative. This can be an extrapolation of the error predicted by a collinear model. The figure makes clear the role played by the saturation physics effects: it will prevent the infrared growth of the collinear parton distribution at high energies. In summary, the trends observed in gluon distribution is the same as those observed in multiplicity-dependent $p_T$ spectra for hard-soft scales \cite{ALICE:2019dfi}, which indicates that the modification of the gluon function is the main source of multiplicity-dependent effects in charged particle spectra.

\begin{figure*}[t]
\includegraphics[width=\linewidth]{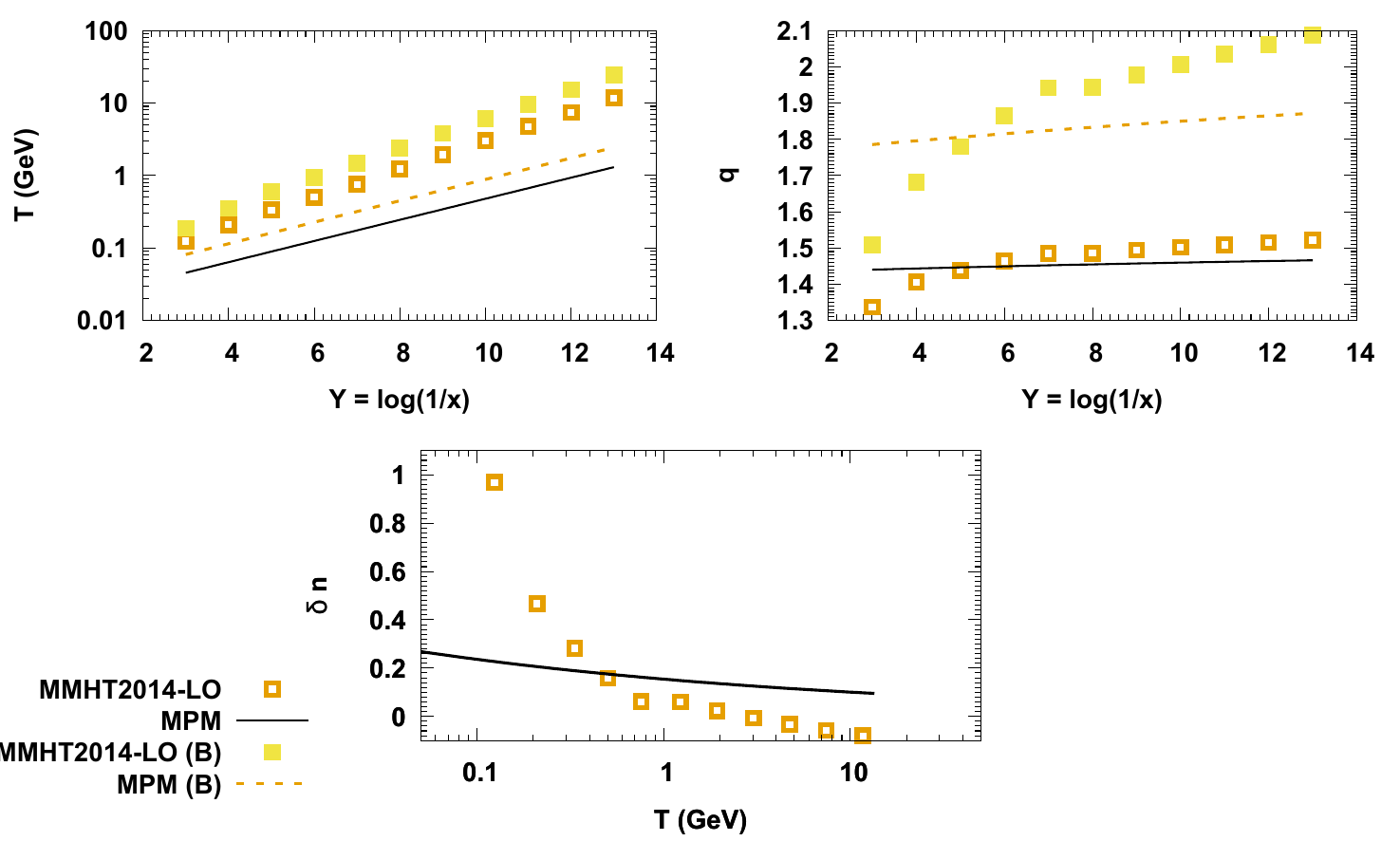}
\caption{Comparison of the UGD from collinear PDF and the MPM scaling curve for the quantities temperature $T$ and non extensive index $q$ as a function of $Y = \ln (1/x)$. The relation between the power index $\delta n$ and the temperature is also presented. The solid square symbols and dashed lines represent the numerical results of thermodynamically consistent Tsallis distribution (see Refs. \cite{Cleymans:2013rfq, Cleymans:2012ya}). }
\label{fig:PARS} 
\end{figure*}

In Figure \ref{fig:PARS} we compare the collinear PDF extrapolated model (open squares) to the MPM scaling curve (solid lines) for the temperature parameters $T$ and the indices $q$ and $\delta n$ as a function of $Y = \ln (1/x)$. These quantities are defined in Eqs.   (\eqref{eq:nepars}-\eqref{eq:pars}). First, the temperature growth is approximated by a power law $T\sim x^{-\lambda}$ as a function of rapidity. However, the predicted behavior from the collinear PDF towards large rapidity is much faster compared to the MPM scaling curve where $\lambda = 0.33$. The nonextensive index $q$ obtained from the relations \eqref{eq:nepars} shows a much faster growth towards its leading order value $3/2$.

In \cite{Cleymans:2013rfq, Cleymans:2012ya}, the authors proposed a thermodynamically consistent Tsallis distribution. It is argued that, in order to reproduce thermodynamic relations, a modification in the identification of the nonextensive parameters, as defined in Eq.~\eqref{eq:nepars}, is necessary.
In this approach, the number of particles is given by the integration over phase space that includes an extra factor of q in the power index.
\begin{equation}
     N = g \sigma_0 \int \frac{d^2k_T}{(2\pi)^2} \left [1 +(q-1) \frac{k_T^2}{T} \right]^{\frac{-q}{q-1}},
 \end{equation}
were $g$ is a degeneracy factor and $\sigma_0$ the proton area. 
In this context, the identification of the nonextensive parameters is modified as
\begin{equation} \label{eq:nepars2}
    q=\frac{2+\delta n}{1 + \delta n}, \qquad T=Q_s^2 (q-1).
\end{equation}
which results in $q \approx 2$ and a higher effective temperature. This possibility is presented in Fig. \ref{fig:PARS} as the label (B) (dashed lines and solid squares). At this point, some comments are in order. The expressions in Eq.~\eqref{eq:nepars} are derived considering the q-Exponential Tsallis distribution, and the results presented in Fig. \ref{fig:PARS}  show the potential impact of choosing alternative descriptions based on non-extensive approaches.  .

The relation between temperature and power index $\delta n$ is also presented for collinear model and MPM parametrization. For $x< 10^{-2}$ ($Y>4$) they are close and the power $\delta n$ decreases slowly starting from $\sim 0.3$ . However, both spectra become harder as the temperature increases. This phenomenon is consistent with that observed in the LHC data for the production of hadrons at high energies; the hadronic spectra become harder as the temperature of the final hadron system increases. The relation between nonextensive parameter $q$ (or $\delta n$) and temperature distinguishes the models and, in principle, this analysis can be made for a large class of gluon distributions in the literature. Therefore, their parameters can be studied under this nonextensive statistical perspective. The relationship between the observable temperature of the final hadron spectrum and the temperature defined here for the gluon system is also to be made.

The $p_T$-spectra of different identified produced hadrons in high energy collisions is very sensitive to the $q$ parameter and the large amount of data available on this observable offers the best system to investigate its behavior as described in next section.

\section{Multiplicity dependent $p_T$ spectra of hadrons}
\label{sec:mult}
Higher multiplicity events at the LHC are often interpreted as evidence of collective behavior. In such events characterized by a large number of charged particles $dN_i/d\eta$ several observables are measured as the enhancement of strangeness \cite{ALICE:2015mpp,CMS:2016zzh,ALICE:2016fzo,Rath:2019cpe}, average transverse momentum, $\langle p_{T_h}\rangle$ \cite{ALICE:2013rdo,ALICE:2019dfi}, correlations \cite{CMS:2023sua} and the production of quarkonium \cite{ALICE:2020msa}. The multiplicity dependence of some observables challenges current models based on QCD. While collectivity and hydrodynamic models are expected to full work on central heavy ion collisions the appearance of this multiplicity dependence in small collision systems is difficult to justify.

Traditional models of perturbative QCD (pQCD) cannot access the bulk properties of the low-$p_T$ spectra, and understanding of soft-hard interface is fundamental. Some models include a multiparton interaction \cite{Ortiz:2020rwg}, where partons can have more than one collision in order to increase multiplicity. Another possibility is to consider gluon saturation dynamics, where strong gluon fields lead to nonlinear effects \cite{Morreale:2021pnn}. In order to understand these new trends presented in the multiplicity dependence data we can employ the scaling analysis \cite{Osada:2019oor,Osada:2020zui,Moriggi:2020zbv,Moriggi:2024tbr,Praszalowicz:2013fsa,Praszalowicz:2015dta}. In this section, we show how the concepts under development in the previous section can be used phenomenologically to understand the multiplicity dependent $p_T$-spectra.

The $p_T$-spectra of produced hadrons are characterized by a power law behavior at high momentum, which can be traced back to the gluon TMDs index $n$ or equivalently to the nonextensive parameter. In the $k_T$ factorization for proton-proton collisions we can write the differential cross section as the convolution of the UGDs from target-projectile protons:
\begin{eqnarray}
\label{eq:fatkt}
E\frac{d^3\sigma}{d\vec{p}^3}^{ab \rightarrow g+X} &=&\frac{\mathcal{A}}{p_T^2}\int d^2k_T\,\phi(x_a,k_T)\phi(x_b , q_T). \\
 &= & f(\tau) \nonumber ,
\end{eqnarray}
where $\tau = p_T^2/Q_s^2$ is a scaling variable and $Q_s(x)$ is the saturation momentum. Concerning  the low-$p_T$ behavior, we follow Ref. \cite{Levin:2010dw} closely where the transverse momentum of the gluon, $k_T$, is replaced by $k_T^2\rightarrow k_T^2+m_{jet}^2$ with $m_{jet}\simeq 0.5\,GeV$ being an effective minijet mass. This procedure naturally regulates the denominator in Eq. \ref{eq:fatkt} due to the presence of a non-zero jet mass (see Ref. \cite{Moriggi:2020zbv} for more details).

There is a dependence on both energy and multiplicity on this quantity, which can be parametrized as an variation of the saturation momentum at each multiplicity class $i$, and the spectra can be calculated form \eqref{eq:fatkt} with the rescaling of the saturation momentum $f(\tau)\rightarrow f(\tau_i)$, where we define 
\begin{equation}
        \tau_i=\frac{Q^2}{\left[X_i Q_{s}(x)\right]^2},.
\end{equation}
 In the above expression, we have considered the variation of the saturation scale in each multiplicity class ($i$) in relation to its minimum bias value, $X_i=Q_{s_i}(x)/ Q_s(x)$.  The energy dependence is, as usual, determined by the $x$-dependent scale. However, there can also be an dependence on the geometry of overlap area and fluctuations as investigated in Ref. \cite{Moriggi:2024tbr}.  The quantity $X_i$ has been obtained from high-multiplicity data at the LHC.

The differential cross section for hadron production can be obtained by considering the overlap area $\langle A_T \rangle$, which is fitted at each multiplicity class,
\begin{equation}\label{eq:Nch}
    E\frac{d^3\sigma_{i}}{d\vec{p}^3}^{ab \rightarrow g+X}\sim \frac{\langle A_T \rangle}{\langle A_{T_{max}} \rangle} f(\tau_i).
\end{equation}

If we assume that the entropy is extensive with respect to the area of interaction, then we have the following relation for low multiplicities, 
\begin{equation}
S_{3/2}(X_i) \sim \langle A_T \rangle \sim (dN_{i}/dy)^{1/3},
\label{eq:prop-AT-S}
\end{equation}
which is in good agreement with experimental data from ALICE \cite{ALICE:2018pal,ALICE:2019dfi}.

\begin{figure*}[t]
\includegraphics[width=\linewidth]{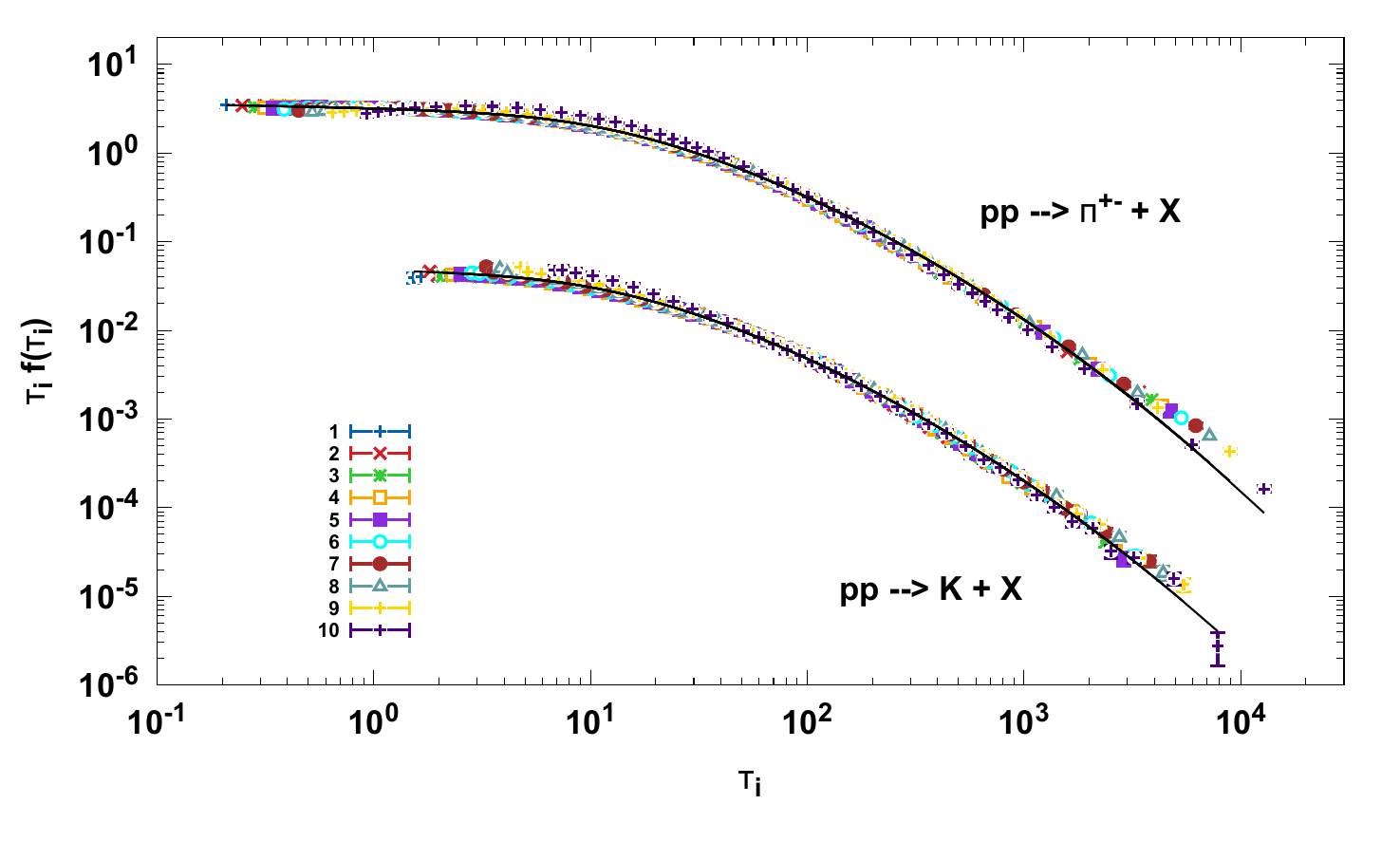}
\caption{Scaling line $\tau_i f(\tau_i)$  for pions and kaons from multiplicity dependent $p_T$-spectra compared to ALICE data \cite{ALICE:2018pal} for several multiplicity classes.}
\label{fig:scaling} 
\end{figure*}

This relation between the overlap area and multiplicity can be used to calculate the multiplicity dependent $p_T$ spectra of different produced final-state hadrons. In Figure \ref{fig:scaling} we show the scaling function $\tau_i f(\tau_i)$ for pions and kaons at $\sqrt{s}=7$ TeV and the comparison with the ALICE data \cite{ALICE:2018pal} for several multiplicity classes. In order to translate $p_T$-spectra from gluon production to identified hadron, we must model the soft hadronization process. In our case a simple local-hadron-parton duality is used as in \cite{Moriggi:2020zbv,Moriggi:2024tbr}. We introduced an effective jet mass $m_j$ and a momentum fraction from hadron to parton $\langle z \rangle$ which can be dependent on the hadronic specie. However, these two quantities are not multiplicity dependent, and all modifications of the spectra are addressed to the initial-state gluon distribution. 
At $\tau_i \sim  10^4$ one starts to see a deviation from the data compatible with the expected scaling violations at high $p_T$.

In this section, it was shown how the rescaling procedure can be used to relate geometric quantities and the multiplicity dependence of the spectrum. In the next section, we show an alternative way of introducing modifications to the multiplicity distribution given by the Glauber multiple scattering formalism in order to include modifications due to the nuclear medium.

\section{Nuclear effects}
\label{sec:DISeA}

The phenomenology of nuclear collisions reveals a range of anomalous effects that have been observed across various experiments over recent decades. Initially, such effects were not anticipated by the parton model of QCD, and their origins continue to be the focus of intense research. It is well understood that the nuclear medium modifies the PDFs of bound nucleons, and that the dynamics of strong interactions can be probed in a distinct regime, where cross sections are amplified due to multiple interactions with the nuclear target.

The precise control of initial cold nuclear matter effects is of fundamental importance to untangle initial/final states nuclear effects due to collectivity in quark-gluon plasma (QGP) phase of heavy-ion collisions. It has been show that part of nuclear shadowing phenomena taken into account in the gluon TMD can change final state distribution parameters in the calculation of azimuthal anisotropy $v_2$ \cite{Moriggi:2023ahi} or nuclear modification factors of high $p_T$ produced pions in $pA$ \cite{Moriggi:2022xbg} and $AA$ collisions \cite{Moriggi:2020qla}. Nuclear modification of TMDs will be of central importance in future electron-ion colliders (EICs) \cite{Accardi:2012qut}.

At small-$x$ gluon distributions are shadowed, i.e. they are reduced in relation to the free per nucleon distribution. Different models have been used to describe its behavior \cite{bCGC,ipsat,Rezaei:2024odh,Krelina:2020ipn,Moriggi:2020qla,NNPDFnuclear,EPPS16,ncteq}. The simplest way to incorporate nuclear modification is to consider the picture of independent interactions between the projectile nucleon and the target with $A$ nucleons. In this model the probability of find $k$ nucleons is given by binomial like distribution and the probability of at least one interaction with target is given by Poisson limit:
\begin{equation}
\begin{aligned}
\label{eq:PNA}
P_{NA}(b) =1-\left( 1-\sigma_{NN}\frac{T_{A}(b)}{A}\right)^A & \approx 1-e^{-\sigma_{NN}T_A(b)}, 
\end{aligned}
\end{equation}
where the approximation implies $\frac{\sigma_{NN}T_{A}(b)}{A} \ll 1$. Even for small nuclei, at $b=0$ one has $T_A(b) < 1 \, \text{fm}^{-1}$. At high energies, we should have $\sigma_{NN} \sim \text{fm}^2$, so this ratio is frequently smaller than unity, which justifies the approximation.

\begin{figure*}[t]
\includegraphics[width=\linewidth]{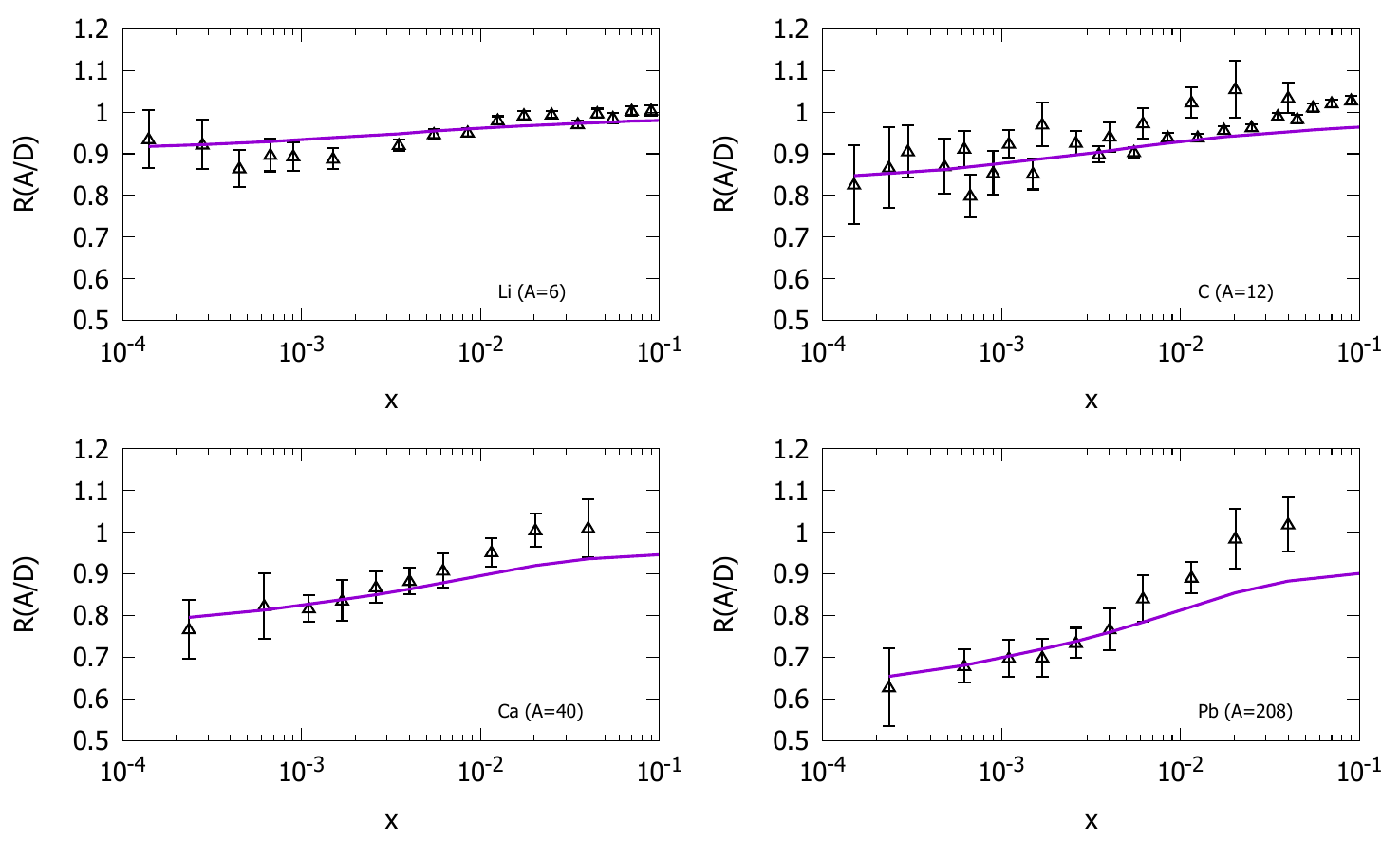}
\caption{Nuclear modification ratios, Eq. \eqref{eq:nuclratios}, as a function of $x$ calculated with multiple scattering formula \eqref{eq:PNA} compared with data from Ref. \cite{Arneodo:1995cs,Adams:1995is}. The results are presented at small-$x$ for the ratios Li, C, Ca and Pb over D.}
\label{fig:DISeAx} 
\end{figure*}

In the color-dipole framework, Mueller \cite{Mueller:1989st} implemented Glauber's multiple-scattering formalism \cite{Glauber:1955qq} to model the dipole’s interaction with a nuclear target. In the target rest frame, the incoming photon dissociates into a quark-antiquark pair. When the probability of interacting with a single nucleon becomes significant, the probability of subsequent scatterings increases, and the dipole then interacts with approximately $A^{1/3}$ nucleons, exchanging two gluons with each.

The probability in momentum space is now dependent of impact parameter and is given by the Fourier transform:
\begin{equation}
\label{eq:fiA}
P_A(x,k_T,b)=\frac{1}{2\pi}\nabla^2_k \, \mathcal{F} \left\{ \frac{1-\Tilde P_{A}(x,r,b)}{r^2}\right\}  .
\end{equation}
One advantage of this expression is that in the dilute regime $k_T \gg Q_s(x)$, it trivially approximates the free nucleon distribution and the nuclear modification factor tends to unity. This limit is important for describing the nuclear modification factors in $pA$ and $AA$ collisions in the region $p_T > Q_s$.

Another advantage of this formalism is that it is independent of extra parameterization of the nuclear distribution, i.e., given the UGD of the proton, we can obtain the nuclear distribution without the need to include any extra dependence on the saturation scale. It should be noted that the distribution applicable to protons also holds for neutrons. For these reasons, we believe that, for the purposes of this work, this is the most appropriate way to obtain the nuclear distribution that will be used in the following analyses.

\begin{figure*}[t]
\includegraphics[width=\linewidth]{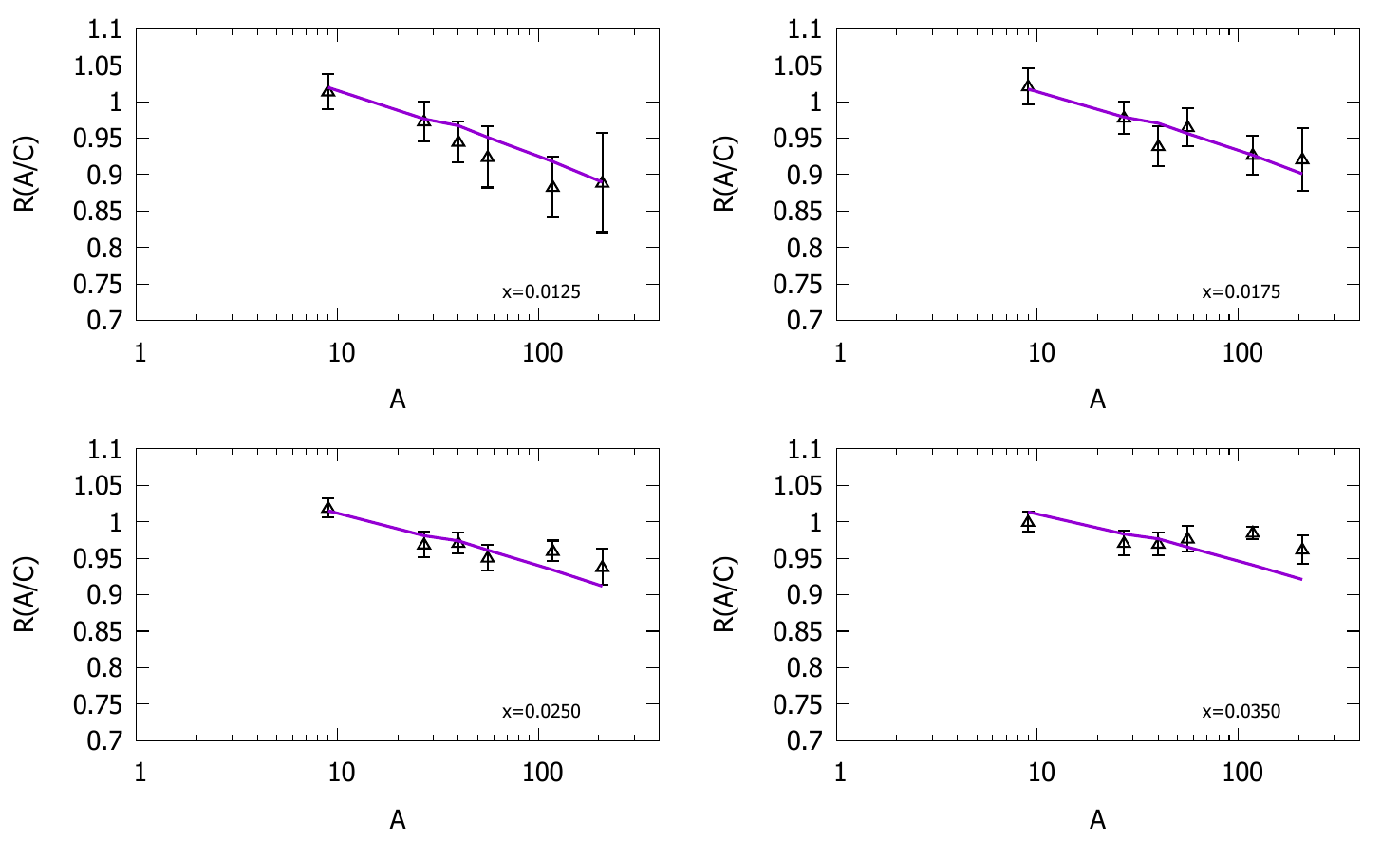}
\caption{Nuclear modification, Eq. \eqref{eq:nuclratios}, as a function of the atomic mass number $A$ for fixed small $x$ compared with data from Ref. \cite{Arneodo:1996ru}. The calculation of the $A$-evolution is compared to the ratios for Be, Al, Ca, Fe, Sn and Pb over C. }
\label{fig:DISeA} 
\end{figure*}

The Glauber-Mueller model was calculated for the $Li,\, Be,\, C, \,Al, \,Ca,\, Fe,\, Sn,\, Au,\, Pb$ nuclei, which will be used in the analysis of nuclear DIS in the next section. In this case, we do not have an analytical form for the $A$ dependence of the distribution, with each nuclear effect calculated from the nuclear density parametrized from elastic $eA$ collision data \cite{DeJager:1974liz, DEVRIES1987495}. For details about the Glauber model and the nuclear overlap function calculations, see, for instance, Refs. \cite{Miller:2007ri,dEnterria:2003xac}.

The nuclear ratio between two different nuclei in DIS off nuclei is defined as:
\begin{equation}
    R(A/B)=\frac{\sigma^{\gamma^*A}(x,Q^2)}{\sigma^{\gamma^*B}(x,Q^2)}\frac{B}{A}.
    \label{eq:nuclratios}
\end{equation}

Figure \ref{fig:DISeAx} shows the ratio for the nuclei $Li,\,C,\,Ca$ and $Pb$ compared to the data in Refs. \cite{Arneodo:1995cs} and \cite{Adams:1995is}. There is a good agreement with the data for all nuclei up to $x<0.05$ where the influence of effects related to anti-shadowing becomes important. Furthermore, we can see that the discrepancy in this limit for light nuclei is much smaller than that for lead. One explanation for this is that suppression of nuclear UGD increases with the atomic number and is therefore larger for lead. In other words, the predicted shadowing for the lead nucleus is larger than expected in the region of higher $x$.

For higher values of $x$, we have a slightly larger suppression than the one observed in the data, since the tin nucleus, like lead, has a large nucleon number, $A=118$. This effect is better presented in Figure \ref{fig:DISeA}, where we compare our results with the data in Ref. \cite{Arneodo:1996ru} as a function of $A$. For $x=0.035$ and $A>100$, the curve slightly underestimates the data.

In summary, the Glauber-Mueller framework effectively captures key nuclear effects such as shadowing and multiple scattering. These findings are essential for exploring how nuclear effects influence hadron production in heavy-ion collisions, as discussed in the next section.

\section{Nuclear modification effects in heavy-ion collisions}
\label{sec:heavyion}

The nuclear modification factor $R_{AA}$ is a key observable in high-energy heavy-ion collisions, used to quantify deviations from the expected results in independent nucleon-nucleon collisions to modifications on a per-nucleon basis. It is defined as:

\begin{equation}
\label{eq:RAA}
R_{AA}(b)=\frac{ \frac{dN_{AA}(b)}{d^2p_{Th}dy}}{ \int d^2s \, T_A(s)T_B(b-s)\frac{d\sigma_{pp}}{d^2p_{Th}dy}},
\end{equation}
where $T_A$ is the nuclear thickness function and $dN_{AA}/d^2p_Tdy$ is the multiplicity distribution on $AA$ collisions.

\begin{figure}[t]
\includegraphics[width=\linewidth]{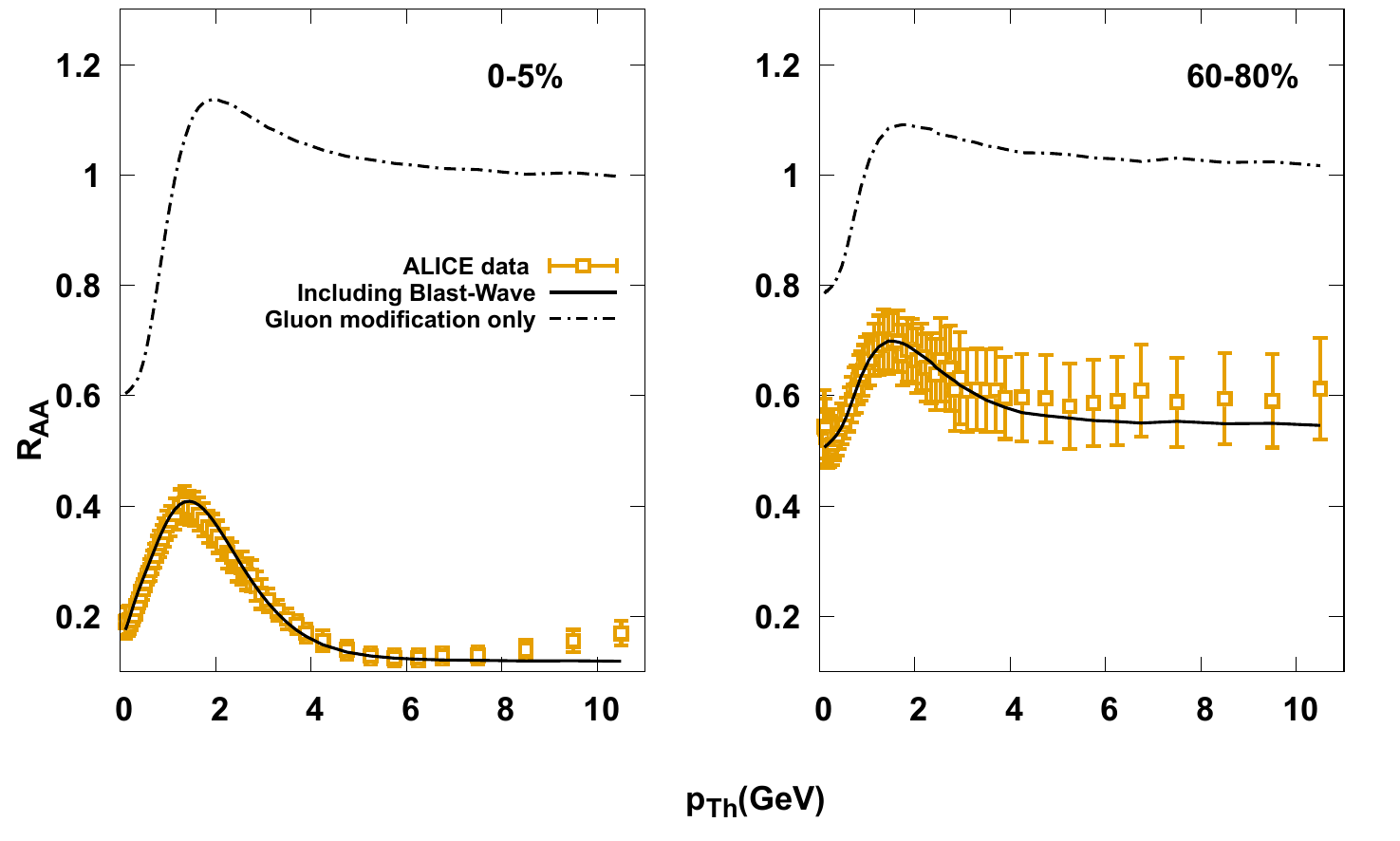}
\caption{Nuclear modification factor $R_{AA}$ for the process $Pb+Pb \rightarrow \pi^{\pm} +X $ as a function of pion transverse momentum at central and peripheral centralities compared with data from ALICE \cite{Adam:2015kca} at 2.76 TeV. The solid lines represent the use of the Glauber-Mueller UGD as the initial particle distribution  in Boltzmann Transport Equation (BTE) in the relaxation time approximation (RTA) and using Blast Wave function as the equilibrium distribution. }
\label{fig:RAA} 
\end{figure}

Several cold nuclear matter effects may contribute to both enhancement and suppression in $R_{AA}$. (i) Modification of PDFs: changes in nuclear PDFs are significant in shaping the final hadron spectrum. For instance, the suppression observed in the nuclear cross-section at low $p_T$ can often be explained by nuclear shadowing. In some works \cite{Helenius:2013bya, Helenius:2015wda}, it is suggested that most nuclear modifications can be effectively described through adjustments in the collinear PDFs. (ii) Multiple scattering effects: since its discovery, the Cronin effect has been understood as an increase in transverse momentum due to multiple scatterings of partons in the proton with the nuclear target. This approach introduces an intrinsic transverse momentum that varies with the centrality of the collision. (iii) Energy loss in medium: Another mechanism contributing to nuclear modification is parton energy loss. Models like those proposed by Vitev et al. \cite{Vitev:2007ve, Kang:2012kc, Arleo:2020eia, Arleo:2020hat} incorporate this effect by assuming a reduction in parton energy through gluon bremsstrahlung induced by interactions with the medium.
    
Although nuclear modifications adequately describe the nuclear structure function observed in deep inelastic scattering (DIS), as shown in \eqref{eq:PNA}, this approach alone is insufficient to capture the full complexity of heavy-ion collisions. To estimate the pion spectrum, we employ the same parametrization as in $pp$ collisions from \eqref{eq:fatkt} with nuclear modification given by \eqref{eq:PNA}. The nuclear modification is given as in the Glauber model from \eqref{eq:RAA}.

Figure \ref{fig:RAA} illustrates the nuclear modification factor in centralities as a function of the transverse momentum $p_{T_h}$ of the charged pions produced at $\sqrt{s}=2.76$ TeV \cite{Adam:2015kca}. We observe a suppression at low $p_{T_h}$ followed by an increase near $Q_s(x)$ resembling the behavior of UGD. Multiple interactions will produce a Cronin-like peak as a result of increasing parton transverse momentum $k_T$. In central collisions where the hot medium modifications are expected to be more important, the suppression is larger than predicted by UGDs modification (dot-dashed lines). On the other hand, in peripheral collisions the difference becomes smaller. To overcome this limitation, see Refs. \cite{Moriggi:2020qla,Moriggi:2023ahi} we use modified UGD calculation as the initial particle distribution function in Boltzmann Transport Equation (BTE) in the relaxation time approximation (RTA).
\begin{eqnarray}
f_{fin}=f_{eq}+\left(f_{in}-f_{eq}  \right)e^{-\frac{t_f}{t_r}},    
\label{eq:finexp}
\end{eqnarray}
A Boltzmann-Gibbs Blast Wave (BGBW) function is taken as an equilibrium function $f_{eq}$ to get the final distribution $f_{fin}$ to describe the particle production in $AA$ collisions. This procedure describes correctly at the same time the $p_T$-spectra, the nuclear modification factors, and the elliptic-flow measurements at the LHC for different centralities. In Fig. \ref{fig:RAA} the result obtained using such a procedure is represented by the solid lines.

In light of nonextensive-like parametrization, we can understand that while temperature changes indicate modifications in the soft component of the spectrum, the high-$p_T$ region retains the same power-law behavior. The nuclear modification factor $R_{AA}$ remains constant and is consistent with the experimental data. This characteristic of power-law parametrization contrasts with exponential-like UGDs, where $R_{AA}$ exhibits exponential growth or suppression.

The process by which the \textit{equilibrium distribution} of the full spectrum develops from its initial configuration remains challenging to fully understand. We hope that investigations like the one presented here can help bridge the gap between the initial partonic configuration and the final hadronic spectrum emerging from QGP-like evolution.

\section{Summary and conclusions}

In summary, this work explored the nonextensive aspects of gluon distributions and their significance in QCD phenomenology. By employing kT factorization and incorporating non-extensive statistical mechanics, we provided new insights into the thermal behavior of gluons and the impact of these distributions in high-energy collisions. Our analysis highlighted how modifications to gluon distributions influence observables across different collision systems, such as $pp$, $pA$, and $AA$ collisions, and emphasized the relevance of scaling behavior in predicting multiplicity-dependent spectra. Future research in this area can further refine our understanding of partonic entropy, improve models of nuclear interactions, and support the development of more precise predictions for hadron collider experiments. These findings contribute to the broader effort to bridge theoretical models with experimental observations, particularly in the context of QGP formation and nuclear modifications in the small-x regime.



\vspace{6pt} 



\authorcontributions{L.S.M. and M.V.T.M. have contributed to the study equally, starting from the conceptualization of the problem, methodology, paper writing and review. All authors have read and agreed to the published version of the manuscript.}

\funding{ This research was funded by the Brazilian National Council for Scientific and Technological  Development (CNPq) under the contract number 303075/2022-8.}

\dataavailability{Not applicable.} 


\conflictsofinterest{: The authors declare no conflict of interest.}


\reftitle{References}


\bibliography{references-MoriggiMachado}

\begin{thebibliography}{999}

\bibitem[Catani et~al.(1990)Catani, Ciafaloni, and Hautmann]{Catani:1990xk}
Catani, S.; Ciafaloni, M.; Hautmann, F.
\newblock {GLUON CONTRIBUTIONS TO SMALL x HEAVY FLAVOR PRODUCTION}.
\newblock {\em Phys. Lett. B} {\bf 1990}, {\em 242},~97--102.
\newblock {\url{https://doi.org/10.1016/0370-2693(90)91601-7}}.

\bibitem[Lipatov et~al.(2014)Lipatov, Lykasov, and Zotov]{Lipatov:2013yra}
Lipatov, A.V.; Lykasov, G.I.; Zotov, N.P.
\newblock {LHC soft physics and transverse momentum dependent gluon density at low x }.
\newblock {\em Phys. Rev. D} {\bf 2014}, {\em 89},~014001,  \href{http://arxiv.org/abs/1310.7893}{{\normalfont [arXiv:hep-ph/1310.7893]}}.
\newblock {\url{https://doi.org/10.1103/PhysRevD.89.014001}}.

\bibitem[Lipatov et~al.(2023)Lipatov, Lykasov, and Malyshev]{Lipatov:2022doa}
Lipatov, A.V.; Lykasov, G.I.; Malyshev, M.A.
\newblock {Toward the global fit of the TMD gluon density in the proton from the LHC data}.
\newblock {\em Phys. Rev. D} {\bf 2023}, {\em 107},~014022,  \href{http://arxiv.org/abs/2211.03727}{{\normalfont [arXiv:hep-ph/2211.03727]}}.
\newblock {\url{https://doi.org/10.1103/PhysRevD.107.014022}}.

\bibitem[Boer(2017)]{Boer:2016bfj}
Boer, D.
\newblock {Gluon TMDs in quarkonium production}.
\newblock {\em Few Body Syst.} {\bf 2017}, {\em 58},~32,  \href{http://arxiv.org/abs/1611.06089}{{\normalfont [arXiv:hep-ph/1611.06089]}}.
\newblock {\url{https://doi.org/10.1007/s00601-016-1198-6}}.

\bibitem[Cisek et~al.(2015)Cisek, Sch\"afer, and Szczurek]{Cisek:2014ala}
Cisek, A.; Sch\"afer, W.; Szczurek, A.
\newblock {Exclusive photoproduction of charmonia in $\gamma p \to V p$ and $p p \to p V p$ reactions within $k_t$-factorization approach}.
\newblock {\em JHEP} {\bf 2015}, {\em 04},~159,  \href{http://arxiv.org/abs/1405.2253}{{\normalfont [arXiv:hep-ph/1405.2253]}}.
\newblock {\url{https://doi.org/10.1007/JHEP04(2015)159}}.

\bibitem[Abdulov et~al.(2021)]{Abdulov:2021ivr}
Abdulov, N.A.;  et~al.
\newblock {TMDlib2 and TMDplotter: a platform for 3D hadron structure studies}.
\newblock {\em Eur. Phys. J. C} {\bf 2021}, {\em 81},~752,  \href{http://arxiv.org/abs/2103.09741}{{\normalfont [arXiv:hep-ph/2103.09741]}}.
\newblock {\url{https://doi.org/10.1140/epjc/s10052-021-09508-8}}.

\bibitem[Moriggi et~al.(2020)Moriggi, Peccini, and Machado]{Moriggi:2020zbv}
Moriggi, L.S.; Peccini, G.M.; Machado, M.V.T.
\newblock {Investigating the inclusive transverse spectra in high-energy $pp$ collisions in the context of geometric scaling framework}.
\newblock {\em Phys. Rev. D} {\bf 2020}, {\em 102},~034016,  \href{http://arxiv.org/abs/2005.07760}{{\normalfont [arXiv:hep-ph/2005.07760]}}.
\newblock {\url{https://doi.org/10.1103/PhysRevD.102.034016}}.

\bibitem[Bacchetta et~al.(2020)Bacchetta, Celiberto, Radici, and Taels]{Bacchetta:2020vty}
Bacchetta, A.; Celiberto, F.G.; Radici, M.; Taels, P.
\newblock {Transverse-momentum-dependent gluon distribution functions in a spectator model}.
\newblock {\em Eur. Phys. J. C} {\bf 2020}, {\em 80},~733,  \href{http://arxiv.org/abs/2005.02288}{{\normalfont [arXiv:hep-ph/2005.02288]}}.
\newblock {\url{https://doi.org/10.1140/epjc/s10052-020-8327-6}}.

\bibitem[Angeles-Martinez et~al.(2015)]{Angeles-Martinez:2015sea}
Angeles-Martinez, R.;  et~al.
\newblock {Transverse Momentum Dependent (TMD) parton distribution functions: status and prospects}.
\newblock {\em Acta Phys. Polon. B} {\bf 2015}, {\em 46},~2501--2534,  \href{http://arxiv.org/abs/1507.05267}{{\normalfont [arXiv:hep-ph/1507.05267]}}.
\newblock {\url{https://doi.org/10.5506/APhysPolB.46.2501}}.

\bibitem[\L{}uszczak et~al.(2022)\L{}uszczak, \L{}uszczak, and Sch\"afer]{Luszczak:2022fkf}
\L{}uszczak, A.; \L{}uszczak, M.; Sch\"afer, W.
\newblock {Unintegrated gluon distributions from the color dipole cross section in the BGK saturation model}.
\newblock {\em Phys. Lett. B} {\bf 2022}, {\em 835},~137582,  \href{http://arxiv.org/abs/2210.02877}{{\normalfont [arXiv:hep-ph/2210.02877]}}.
\newblock {\url{https://doi.org/10.1016/j.physletb.2022.137582}}.

\bibitem[Golec-Biernat and Sapeta(2018)]{GBWnovo}
Golec-Biernat, K.; Sapeta, S.
\newblock {Saturation model of DIS : an update}.
\newblock {\em JHEP} {\bf 2018}, {\em 03},~102,  \href{http://arxiv.org/abs/1711.11360}{{\normalfont [arXiv:hep-ph/1711.11360]}}.
\newblock {\url{https://doi.org/10.1007/JHEP03(2018)102}}.

\bibitem[Kutak and Sapeta(2012)]{Kutak:2012rf}
Kutak, K.; Sapeta, S.
\newblock {Gluon saturation in dijet production in p-Pb collisions at Large Hadron Collider}.
\newblock {\em Phys. Rev. D} {\bf 2012}, {\em 86},~094043,  \href{http://arxiv.org/abs/1205.5035}{{\normalfont [arXiv:hep-ph/1205.5035]}}.
\newblock {\url{https://doi.org/10.1103/PhysRevD.86.094043}}.

\bibitem[Moriggi et~al.(2024)Moriggi, Ramos, and Machado]{Moriggi:2024tbr}
Moriggi, L.S.; Ramos, G.S.; Machado, M.V.T.
\newblock {Multiplicity dependence of the pT-spectra for identified particles and its relationship with partonic entropy}.
\newblock {\em Phys. Rev. D} {\bf 2024}, {\em 110},~034005,  \href{http://arxiv.org/abs/2405.01712}{{\normalfont [arXiv:hep-ph/2405.01712]}}.
\newblock {\url{https://doi.org/10.1103/PhysRevD.110.034005}}.

\bibitem[McLerran and Praszalowicz(2015)]{McLerran:2014apa}
McLerran, L.; Praszalowicz, M.
\newblock {Geometrical Scaling and the Dependence of the Average Transverse Momentum on the Multiplicity and Energy for the ALICE Experiment}.
\newblock {\em Phys. Lett. B} {\bf 2015}, {\em 741},~246--251,  \href{http://arxiv.org/abs/1407.6687}{{\normalfont [arXiv:hep-ph/1407.6687]}}.
\newblock {\url{https://doi.org/10.1016/j.physletb.2014.12.046}}.

\bibitem[Stasto et~al.(2001)Stasto, Golec-Biernat, and Kwiecinski]{Stasto}
Stasto, A.M.; Golec-Biernat, K.J.; Kwiecinski, J.
\newblock {Geometric scaling for the total gamma* p cross-section in the low x region}.
\newblock {\em Phys. Rev. Lett.} {\bf 2001}, {\em 86},~596--599,  \href{http://arxiv.org/abs/hep-ph/0007192}{{\normalfont [arXiv:hep-ph/hep-ph/0007192]}}.
\newblock {\url{https://doi.org/10.1103/PhysRevLett.86.596}}.

\bibitem[Praszalowicz(2011)]{Praszalowicz:2011tc}
Praszalowicz, M.
\newblock {Improved Geometrical Scaling at the LHC}.
\newblock {\em Phys. Rev. Lett.} {\bf 2011}, {\em 106},~142002,  \href{http://arxiv.org/abs/1101.0585}{{\normalfont [arXiv:hep-ph/1101.0585]}}.
\newblock {\url{https://doi.org/10.1103/PhysRevLett.106.142002}}.

\bibitem[Praszalowicz(2013)]{Praszalowicz:2013fsa}
Praszalowicz, M.
\newblock {Geometrical scaling for identified particles}.
\newblock {\em Phys. Lett. B} {\bf 2013}, {\em 727},~461--467,  \href{http://arxiv.org/abs/1308.5911}{{\normalfont [arXiv:hep-ph/1308.5911]}}.
\newblock {\url{https://doi.org/10.1016/j.physletb.2013.10.067}}.

\bibitem[Prasza\l~owicz and Francuz(2015)]{Praszalowicz:2015dta}
Prasza\l~owicz, M.; Francuz, A.
\newblock {Geometrical Scaling in Inelastic Inclusive Particle Production at the LHC}.
\newblock {\em Phys. Rev. D} {\bf 2015}, {\em 92},~074036,  \href{http://arxiv.org/abs/1507.08186}{{\normalfont [arXiv:hep-ph/1507.08186]}}.
\newblock {\url{https://doi.org/10.1103/PhysRevD.92.074036}}.

\bibitem[Osada and Kumaoka(2019)]{Osada:2019oor}
Osada, T.; Kumaoka, T.
\newblock {Saturation momentum scale extracted from semi-inclusive transverse spectra in high-energy pp collisions}.
\newblock {\em Phys. Rev. C} {\bf 2019}, {\em 100},~034906,  \href{http://arxiv.org/abs/1904.10823}{{\normalfont [arXiv:hep-ph/1904.10823]}}.
\newblock {\url{https://doi.org/10.1103/PhysRevC.100.034906}}.

\bibitem[Osada(2021)]{Osada:2020zui}
Osada, T.
\newblock {Multiplicity-dependent saturation momentum in $p$-Pb collisions at 5.02 TeV}.
\newblock {\em Phys. Rev. C} {\bf 2021}, {\em 103},~024911,  \href{http://arxiv.org/abs/2011.00456}{{\normalfont [arXiv:nucl-th/2011.00456]}}.
\newblock {\url{https://doi.org/10.1103/PhysRevC.103.024911}}.

\bibitem[Peccini et~al.(2020)Peccini, Moriggi, and Machado]{Peccini:2020jkj}
Peccini, G.M.; Moriggi, L.S.; Machado, M.V.T.
\newblock {Dilepton production through timelike Compton scattering within the $k_T$ -factorization approach}.
\newblock {\em Phys. Rev. D} {\bf 2020}, {\em 102},~094015,  \href{http://arxiv.org/abs/2010.03101}{{\normalfont [arXiv:hep-ph/2010.03101]}}.
\newblock {\url{https://doi.org/10.1103/PhysRevD.102.094015}}.

\bibitem[Peccini et~al.(2021)Peccini, Moriggi, and Machado]{Peccini:2021rbt}
Peccini, G.M.; Moriggi, L.S.; Machado, M.V.T.
\newblock {Exclusive dilepton production via timelike Compton scattering in heavy ion collisions}.
\newblock {\em Phys. Rev. D} {\bf 2021}, {\em 103},~054009,  \href{http://arxiv.org/abs/2101.08338}{{\normalfont [arXiv:hep-ph/2101.08338]}}.
\newblock {\url{https://doi.org/10.1103/PhysRevD.103.054009}}.

\bibitem[Cisek et~al.(2023)Cisek, Sch\"afer, and Szczurek]{Cisek:2022yjj}
Cisek, A.; Sch\"afer, W.; Szczurek, A.
\newblock {Exclusive production of \ensuremath{\rho} meson in gamma-proton collisions: d\ensuremath{\sigma}/dt and the role of helicity flip processes}.
\newblock {\em Phys. Lett. B} {\bf 2023}, {\em 836},~137595,  \href{http://arxiv.org/abs/2209.06578}{{\normalfont [arXiv:hep-ph/2209.06578]}}.
\newblock {\url{https://doi.org/10.1016/j.physletb.2022.137595}}.

\bibitem[Sampaio~dos Santos et~al.(2023)Sampaio~dos Santos, Gil~da Silveira, and Machado]{SampaiodosSantos:2021tfh}
Sampaio~dos Santos, G.; Gil~da Silveira, G.; Machado, M.V.T.
\newblock {Charmed meson production based on dipole transverse momentum representation in high energy hadron-hadron collisions available at the LHC}.
\newblock {\em Phys. Lett. B} {\bf 2023}, {\em 838},~137667,  \href{http://arxiv.org/abs/2108.01562}{{\normalfont [arXiv:hep-ph/2108.01562]}}.
\newblock {\url{https://doi.org/10.1016/j.physletb.2023.137667}}.

\bibitem[Santos et~al.(2023)Santos, da~Silveira, and Machado]{Santos:2023yep}
Santos, G.S.d.; da~Silveira, G.G.; Machado, M.V.T.
\newblock {Double charmed meson in pp and pA collisions production at the LHC within the dipole approach in momentum representation}.
\newblock {\em Eur. Phys. J. C} {\bf 2023}, {\em 83},~862,  \href{http://arxiv.org/abs/2309.07686}{{\normalfont [arXiv:hep-ph/2309.07686]}}.
\newblock {\url{https://doi.org/10.1140/epjc/s10052-023-12028-2}}.

\bibitem[Moriggi et~al.(2021)Moriggi, Peccini, and Machado]{Moriggi:2020qla}
Moriggi, L.S.; Peccini, G.M.; Machado, M.V.T.
\newblock {Role of nuclear gluon distribution on particle production in heavy ion collisions}.
\newblock {\em Phys. Rev. D} {\bf 2021}, {\em 103},~034025,  \href{http://arxiv.org/abs/2012.05388}{{\normalfont [arXiv:hep-ph/2012.05388]}}.
\newblock {\url{https://doi.org/10.1103/PhysRevD.103.034025}}.

\bibitem[Moriggi et~al.(2023)Moriggi, dos Santos~Rocha, and Machado]{Moriggi:2023ahi}
Moriggi, L.S.; dos Santos~Rocha, E.; Machado, M.V.T.
\newblock {Study of the azimuthal asymmetry in heavy ion collisions combining initial state momentum orientation and final state collective effects}.
\newblock {\em Phys. Rev. D} {\bf 2023}, {\em 108},~074013,  \href{http://arxiv.org/abs/2308.00181}{{\normalfont [arXiv:hep-ph/2308.00181]}}.
\newblock {\url{https://doi.org/10.1103/PhysRevD.108.074013}}.

\bibitem[Hagedorn(1983)]{Hagedorn}
Hagedorn, R.
\newblock {Multiplicities, $p_T$ Distributions and the Expected Hadron $\to$ Quark - Gluon Phase Transition}.
\newblock {\em Riv. Nuovo Cim.} {\bf 1983}, {\em 6N10},~1--50.
\newblock {\url{https://doi.org/10.1007/BF02740917}}.

\bibitem[Wong et~al.(2015)Wong, Wilk, Cirto, and Tsallis]{Wong:2015mba}
Wong, C.Y.; Wilk, G.; Cirto, L.J.L.; Tsallis, C.
\newblock {From QCD-based hard-scattering to nonextensive statistical mechanical descriptions of transverse momentum spectra in high-energy $pp$ and $p\bar p$ collisions}.
\newblock {\em Phys. Rev. D} {\bf 2015}, {\em 91},~114027,  \href{http://arxiv.org/abs/1505.02022}{{\normalfont [arXiv:hep-ph/1505.02022]}}.
\newblock {\url{https://doi.org/10.1103/PhysRevD.91.114027}}.

\bibitem[Bhattacharyya et~al.(2018)Bhattacharyya, Cleymans, Mogliacci, Parvan, Sorin, and Teryaev]{Bhattacharyya:2017cdk}
Bhattacharyya, T.; Cleymans, J.; Mogliacci, S.; Parvan, A.S.; Sorin, A.S.; Teryaev, O.V.
\newblock {Non-extensivity of the QCD p$_{T}$-spectra}.
\newblock {\em Eur. Phys. J. A} {\bf 2018}, {\em 54},~222,  \href{http://arxiv.org/abs/1712.08334}{{\normalfont [arXiv:hep-ph/1712.08334]}}.
\newblock {\url{https://doi.org/10.1140/epja/i2018-12647-6}}.

\bibitem[B\'\i{}r\'o et~al.(2017)B\'\i{}r\'o, Barnaf\"oldi, Bir\'o, \"Urm\"ossy, and Tak\'acs]{Biro:2017arf}
B\'\i{}r\'o, G.; Barnaf\"oldi, G.G.; Bir\'o, T.S.; \"Urm\"ossy, K.; Tak\'acs, A.
\newblock {Systematic Analysis of the Non-extensive Statistical Approach in High Energy Particle Collisions - Experiment vs. Theory}.
\newblock {\em Entropy} {\bf 2017}, {\em 19},~88,  \href{http://arxiv.org/abs/1702.02842}{{\normalfont [arXiv:hep-ph/1702.02842]}}.
\newblock {\url{https://doi.org/10.3390/e19030088}}.

\bibitem[B\'\i{}r\'o et~al.(2020)B\'\i{}r\'o, Barnaf\"oldi, and Bir\'o]{Biro:2020kve}
B\'\i{}r\'o, G.; Barnaf\"oldi, G.G.; Bir\'o, T.S.
\newblock {Tsallis-thermometer: a QGP indicator for large and small collisional systems}.
\newblock {\em J. Phys. G} {\bf 2020}, {\em 47},~105002,  \href{http://arxiv.org/abs/2003.03278}{{\normalfont [arXiv:hep-ph/2003.03278]}}.
\newblock {\url{https://doi.org/10.1088/1361-6471/ab8dcb}}.

\bibitem[Akhil and Tiwari(2024)]{Akhil:2023xpb}
Akhil, A.; Tiwari, S.K.
\newblock {Exploring anisotropic flow via the Boltzmann transport equation employing the Tsallis blast wave description at LHC energies}.
\newblock {\em J. Phys. G} {\bf 2024}, {\em 51},~035002,  \href{http://arxiv.org/abs/2309.06128}{{\normalfont [arXiv:hep-ph/2309.06128]}}.
\newblock {\url{https://doi.org/10.1088/1361-6471/ad1eb9}}.

\bibitem[Li et~al.(2021)Li, Liu, and Olimov]{Li:2020lww}
Li, L.L.; Liu, F.H.; Olimov, K.K.
\newblock {Excitation Functions of Tsallis-Like Parameters in High-Energy Nucleus\textendash{}Nucleus Collisions}.
\newblock {\em Entropy} {\bf 2021}, {\em 23},~478,  \href{http://arxiv.org/abs/2006.15333}{{\normalfont [arXiv:hep-ph/2006.15333]}}.
\newblock {\url{https://doi.org/10.3390/e23040478}}.

\bibitem[Sharma et~al.(2019)Sharma, Cleymans, Hippolyte, and Paradza]{Sharma:2018jqf}
Sharma, N.; Cleymans, J.; Hippolyte, B.; Paradza, M.
\newblock {A Comparison of p-p, p-Pb, Pb-Pb Collisions in the Thermal Model: Multiplicity Dependence of Thermal Parameters}.
\newblock {\em Phys. Rev. C} {\bf 2019}, {\em 99},~044914,  \href{http://arxiv.org/abs/1811.00399}{{\normalfont [arXiv:hep-ph/1811.00399]}}.
\newblock {\url{https://doi.org/10.1103/PhysRevC.99.044914}}.

\bibitem[Khuntia et~al.(2017)Khuntia, Tripathy, Sahoo, and Cleymans]{Khuntia:2017ite}
Khuntia, A.; Tripathy, S.; Sahoo, R.; Cleymans, J.
\newblock {Multiplicity Dependence of Non-extensive Parameters for Strange and Multi-Strange Particles in Proton-Proton Collisions at $\sqrt{s}= 7$ TeV at the LHC}.
\newblock {\em Eur. Phys. J. A} {\bf 2017}, {\em 53},~103,  \href{http://arxiv.org/abs/1702.06885}{{\normalfont [arXiv:hep-ph/1702.06885]}}.
\newblock {\url{https://doi.org/10.1140/epja/i2017-12291-8}}.

\bibitem[Bhattacharyya et~al.(2016)Bhattacharyya, Cleymans, and Mogliacci]{Bhattacharyya:2016lrk}
Bhattacharyya, T.; Cleymans, J.; Mogliacci, S.
\newblock {Analytic results for the Tsallis thermodynamic variables}.
\newblock {\em Phys. Rev. D} {\bf 2016}, {\em 94},~094026,  \href{http://arxiv.org/abs/1608.08965}{{\normalfont [arXiv:cond-mat.stat-mech/1608.08965]}}.
\newblock {\url{https://doi.org/10.1103/PhysRevD.94.094026}}.

\bibitem[Parvan et~al.(2017)Parvan, Teryaev, and Cleymans]{Parvan:2016rln}
Parvan, A.S.; Teryaev, O.V.; Cleymans, J.
\newblock {Systematic Comparison of Tsallis Statistics for Charged Pions Produced In $pp$ Collisions}.
\newblock {\em Eur. Phys. J. A} {\bf 2017}, {\em 53},~102,  \href{http://arxiv.org/abs/1607.01956}{{\normalfont [arXiv:nucl-th/1607.01956]}}.
\newblock {\url{https://doi.org/10.1140/epja/i2017-12301-y}}.

\bibitem[Tripathy et~al.(2016)Tripathy, Bhattacharyya, Garg, Kumar, Sahoo, and Cleymans]{Tripathy:2016hlg}
Tripathy, S.; Bhattacharyya, T.; Garg, P.; Kumar, P.; Sahoo, R.; Cleymans, J.
\newblock {Nuclear Modification Factor Using Tsallis Non-extensive Statistics}.
\newblock {\em Eur. Phys. J. A} {\bf 2016}, {\em 52},~289,  \href{http://arxiv.org/abs/1606.06898}{{\normalfont [arXiv:nucl-th/1606.06898]}}.
\newblock {\url{https://doi.org/10.1140/epja/i2016-16289-4}}.

\bibitem[Han et~al.(2020)Han, Xing, Wang, Fu, Wang, and Chen]{Han:2018wsw}
Han, C.; Xing, H.; Wang, X.; Fu, Q.; Wang, R.; Chen, X.
\newblock {Pion Valence Quark Distributions from Maximum Entropy Method}.
\newblock {\em Phys. Lett. B} {\bf 2020}, {\em 800},~135066,  \href{http://arxiv.org/abs/1809.01549}{{\normalfont [arXiv:hep-ph/1809.01549]}}.
\newblock {\url{https://doi.org/10.1016/j.physletb.2019.135066}}.

\bibitem[Chen et~al.(2024)Chen, Wang, Cai, Chen, and Wang]{Chen:2024dhz}
Chen, J.; Wang, X.; Cai, Y.; Chen, X.; Wang, Q.
\newblock {Valence Quark Distributions in Pions: Insights from Tsallis Entropy} {\bf 2024}.
\newblock  \href{http://arxiv.org/abs/2408.03068}{{\normalfont [arXiv:hep-ph/2408.03068]}}.

\bibitem[Gribov et~al.(1983)Gribov, Levin, and Ryskin]{GLR}
Gribov, L.V.; Levin, E.M.; Ryskin, M.G.
\newblock {Semihard Processes in QCD}.
\newblock {\em Phys. Rept.} {\bf 1983}, {\em 100},~1--150.
\newblock {\url{https://doi.org/10.1016/0370-1573(83)90022-4}}.

\bibitem[Mueller and Qiu(1986)]{MQ}
Mueller, A.H.; Qiu, J.
\newblock Gluon recombination and shadowing at small values of x.
\newblock {\em Nuclear Physics B} {\bf 1986}, {\em 268},~427--452.

\bibitem[Kharzeev and Levin(2017)]{Kharzeev:2017qzs}
Kharzeev, D.E.; Levin, E.M.
\newblock {Deep inelastic scattering as a probe of entanglement}.
\newblock {\em Phys. Rev. D} {\bf 2017}, {\em 95},~114008,  \href{http://arxiv.org/abs/1702.03489}{{\normalfont [arXiv:hep-ph/1702.03489]}}.
\newblock {\url{https://doi.org/10.1103/PhysRevD.95.114008}}.

\bibitem[Hentschinski and Kutak(2022)]{Hentschinski:2021aux}
Hentschinski, M.; Kutak, K.
\newblock {Evidence for the maximally entangled low x proton in Deep Inelastic Scattering from H1 data}.
\newblock {\em Eur. Phys. J. C} {\bf 2022}, {\em 82},~111,  \href{http://arxiv.org/abs/2110.06156}{{\normalfont [arXiv:hep-ph/2110.06156]}}.
\newblock [Erratum: Eur.Phys.J.C 83, 1147 (2023)], {\url{https://doi.org/10.1140/epjc/s10052-022-10056-y}}.

\bibitem[Hentschinski et~al.(2023)Hentschinski, Kharzeev, Kutak, and Tu]{Hentschinski:2023izh}
Hentschinski, M.; Kharzeev, D.E.; Kutak, K.; Tu, Z.
\newblock {Probing the Onset of Maximal Entanglement inside the Proton in Diffractive Deep Inelastic Scattering}.
\newblock {\em Phys. Rev. Lett.} {\bf 2023}, {\em 131},~241901,  \href{http://arxiv.org/abs/2305.03069}{{\normalfont [arXiv:hep-ph/2305.03069]}}.
\newblock {\url{https://doi.org/10.1103/PhysRevLett.131.241901}}.

\bibitem[Hentschinski et~al.(2024)Hentschinski, Kharzeev, Kutak, and Tu]{Hentschinski:2024gaa}
Hentschinski, M.; Kharzeev, D.E.; Kutak, K.; Tu, Z.
\newblock {QCD evolution of entanglement entropy} {\bf 2024}.
\newblock  \href{http://arxiv.org/abs/2408.01259}{{\normalfont [arXiv:hep-ph/2408.01259]}}.

\bibitem[Harland-Lang et~al.(2015)Harland-Lang, Martin, Motylinski, and Thorne]{MMHT2014}
Harland-Lang, L.A.; Martin, A.D.; Motylinski, P.; Thorne, R.S.
\newblock {Parton distributions in the LHC era: MMHT 2014 PDFs}.
\newblock {\em Eur. Phys. J.} {\bf 2015}, {\em C75},~204,  \href{http://arxiv.org/abs/1412.3989}{{\normalfont [arXiv:hep-ph/1412.3989]}}.
\newblock {\url{https://doi.org/10.1140/epjc/s10052-015-3397-6}}.

\bibitem[Hou et~al.(2021)]{Hou:2019efy}
Hou, T.J.;  et~al.
\newblock {New CTEQ global analysis of quantum chromodynamics with high-precision data from the LHC}.
\newblock {\em Phys. Rev. D} {\bf 2021}, {\em 103},~014013,  \href{http://arxiv.org/abs/1912.10053}{{\normalfont [arXiv:hep-ph/1912.10053]}}.
\newblock {\url{https://doi.org/10.1103/PhysRevD.103.014013}}.

\bibitem[Bailey et~al.(2021)Bailey, Cridge, Harland-Lang, Martin, and Thorne]{Bailey:2020ooq}
Bailey, S.; Cridge, T.; Harland-Lang, L.A.; Martin, A.D.; Thorne, R.S.
\newblock {Parton distributions from LHC, HERA, Tevatron and fixed target data: MSHT20 PDFs}.
\newblock {\em Eur. Phys. J. C} {\bf 2021}, {\em 81},~341,  \href{http://arxiv.org/abs/2012.04684}{{\normalfont [arXiv:hep-ph/2012.04684]}}.
\newblock {\url{https://doi.org/10.1140/epjc/s10052-021-09057-0}}.

\bibitem[Ball et~al.(2017)]{NNPDF:2017mvq}
Ball, R.D.;  et~al.
\newblock {Parton distributions from high-precision collider data}.
\newblock {\em Eur. Phys. J. C} {\bf 2017}, {\em 77},~663,  \href{http://arxiv.org/abs/1706.00428}{{\normalfont [arXiv:hep-ph/1706.00428]}}.
\newblock {\url{https://doi.org/10.1140/epjc/s10052-017-5199-5}}.

\bibitem[Abramowicz et~al.(2015)]{HERAPDF}
Abramowicz, H.;  et~al.
\newblock {Combination of measurements of inclusive deep inelastic ${e^{\pm }p}$ scattering cross sections and QCD analysis of HERA data}.
\newblock {\em Eur. Phys. J.} {\bf 2015}, {\em C75},~580,  \href{http://arxiv.org/abs/1506.06042}{{\normalfont [arXiv:hep-ex/1506.06042]}}.
\newblock {\url{https://doi.org/10.1140/epjc/s10052-015-3710-4}}.

\bibitem[Dokshitzer(1977)]{D}
Dokshitzer, Y.L.
\newblock {Calculation of the Structure Functions for Deep Inelastic Scattering and e+ e- Annihilation by Perturbation Theory in Quantum Chromodynamics.}
\newblock {\em Sov. Phys. JETP} {\bf 1977}, {\em 46},~641--653.
\newblock [Zh. Eksp. Teor. Fiz.73,1216(1977)].

\bibitem[Gribov and Lipatov(1972)]{GL}
Gribov, V.N.; Lipatov, L.N.
\newblock {Deep inelastic e p scattering in perturbation theory}.
\newblock {\em Sov. J. Nucl. Phys.} {\bf 1972}, {\em 15},~438--450.
\newblock [Yad. Fiz.15,781(1972)].

\bibitem[Altarelli and Parisi(1977)]{AP}
Altarelli, G.; Parisi, G.
\newblock {Asymptotic Freedom in Parton Language}.
\newblock {\em Nucl. Phys.} {\bf 1977}, {\em B126},~298--318.
\newblock {\url{https://doi.org/10.1016/0550-3213(77)90384-4}}.

\bibitem[Kuraev et~al.(1977)Kuraev, Lipatov, and Fadin]{BFKL1}
Kuraev, E.A.; Lipatov, L.N.; Fadin, V.S.
\newblock {The Pomeranchuk Singularity in Nonabelian Gauge Theories}.
\newblock {\em Sov. Phys. JETP} {\bf 1977}, {\em 45},~199--204.
\newblock [Zh. Eksp. Teor. Fiz.72,377(1977)].

\bibitem[Balitsky and Lipatov(1978)]{BFKL2}
Balitsky, I.I.; Lipatov, L.N.
\newblock {The Pomeranchuk Singularity in Quantum Chromodynamics}.
\newblock {\em Sov. J. Nucl. Phys.} {\bf 1978}, {\em 28},~822--829.
\newblock [Yad. Fiz.28,1597(1978)].

\bibitem[Balitsky(1996)]{BK1}
Balitsky, I.
\newblock {Operator expansion for high-energy scattering}.
\newblock {\em Nucl. Phys.} {\bf 1996}, {\em B463},~99--160,  \href{http://arxiv.org/abs/hep-ph/9509348}{{\normalfont [arXiv:hep-ph/hep-ph/9509348]}}.
\newblock {\url{https://doi.org/10.1016/0550-3213(95)00638-9}}.

\bibitem[Kovchegov(1999)]{BK2}
Kovchegov, Y.V.
\newblock {Small x F(2) structure function of a nucleus including multiple pomeron exchanges}.
\newblock {\em Phys. Rev.} {\bf 1999}, {\em D60},~034008,  \href{http://arxiv.org/abs/hep-ph/9901281}{{\normalfont [arXiv:hep-ph/hep-ph/9901281]}}.
\newblock {\url{https://doi.org/10.1103/PhysRevD.60.034008}}.

\bibitem[Tsallis(1988)]{Tsallis:1987eu}
Tsallis, C.
\newblock {Possible Generalization of Boltzmann-Gibbs Statistics}.
\newblock {\em J. Statist. Phys.} {\bf 1988}, {\em 52},~479--487.
\newblock {\url{https://doi.org/10.1007/BF01016429}}.

\bibitem[Tsallis et~al.(1998)Tsallis, Mendes, and Plastino]{Tsallis:1998ws}
Tsallis, C.; Mendes, R.S.; Plastino, A.R.
\newblock {The Role of constraints within generalized nonextensive statistics}.
\newblock {\em Physica A} {\bf 1998}, {\em 261},~534.
\newblock {\url{https://doi.org/10.1016/S0378-4371(98)00437-3}}.

\bibitem[Prato and Tsallis(1999)]{PhysRevE.60.2398}
Prato, D.; Tsallis, C.
\newblock Nonextensive foundation of L\'evy distributions.
\newblock {\em Phys. Rev. E} {\bf 1999}, {\em 60},~2398--2401.
\newblock {\url{https://doi.org/10.1103/PhysRevE.60.2398}}.

\bibitem[Zanette and Alemany(1995)]{PhysRevLett.75.366}
Zanette, D.H.; Alemany, P.A.
\newblock Thermodynamics of Anomalous Diffusion.
\newblock {\em Phys. Rev. Lett.} {\bf 1995}, {\em 75},~366--369.
\newblock {\url{https://doi.org/10.1103/PhysRevLett.75.366}}.

\bibitem[Acharya et~al.(2019)]{ALICE:2019dfi}
Acharya, S.;  et~al.
\newblock {Charged-particle production as a function of multiplicity and transverse spherocity in pp collisions at $\sqrt{s} =5.02$ and 13 TeV}.
\newblock {\em Eur. Phys. J. C} {\bf 2019}, {\em 79},~857,  \href{http://arxiv.org/abs/1905.07208}{{\normalfont [arXiv:nucl-ex/1905.07208]}}.
\newblock {\url{https://doi.org/10.1140/epjc/s10052-019-7350-y}}.

\bibitem[Cleymans et~al.(2013)Cleymans, Lykasov, Parvan, Sorin, Teryaev, and Worku]{Cleymans:2013rfq}
Cleymans, J.; Lykasov, G.I.; Parvan, A.S.; Sorin, A.S.; Teryaev, O.V.; Worku, D.
\newblock {Systematic properties of the Tsallis Distribution: Energy Dependence of Parameters in High-Energy p-p Collisions}.
\newblock {\em Phys. Lett. B} {\bf 2013}, {\em 723},~351--354,  \href{http://arxiv.org/abs/1302.1970}{{\normalfont [arXiv:hep-ph/1302.1970]}}.
\newblock {\url{https://doi.org/10.1016/j.physletb.2013.05.029}}.

\bibitem[Cleymans and Worku(2012)]{Cleymans:2012ya}
Cleymans, J.; Worku, D.
\newblock {Relativistic Thermodynamics: Transverse Momentum Distributions in High-Energy Physics}.
\newblock {\em Eur. Phys. J. A} {\bf 2012}, {\em 48},~160,  \href{http://arxiv.org/abs/1203.4343}{{\normalfont [arXiv:hep-ph/1203.4343]}}.
\newblock {\url{https://doi.org/10.1140/epja/i2012-12160-0}}.

\bibitem[Adam et~al.(2016)]{ALICE:2015mpp}
Adam, J.;  et~al.
\newblock {Multi-strange baryon production in p-Pb collisions at $\sqrt{s_\mathbf{NN}}=5.02$ TeV}.
\newblock {\em Phys. Lett. B} {\bf 2016}, {\em 758},~389--401,  \href{http://arxiv.org/abs/1512.07227}{{\normalfont [arXiv:nucl-ex/1512.07227]}}.
\newblock {\url{https://doi.org/10.1016/j.physletb.2016.05.027}}.

\bibitem[Khachatryan et~al.(2017)]{CMS:2016zzh}
Khachatryan, V.;  et~al.
\newblock {Multiplicity and rapidity dependence of strange hadron production in pp, pPb, and PbPb collisions at the LHC}.
\newblock {\em Phys. Lett. B} {\bf 2017}, {\em 768},~103--129,  \href{http://arxiv.org/abs/1605.06699}{{\normalfont [arXiv:nucl-ex/1605.06699]}}.
\newblock {\url{https://doi.org/10.1016/j.physletb.2017.01.075}}.

\bibitem[Adam et~al.(2017)]{ALICE:2016fzo}
Adam, J.;  et~al.
\newblock {Enhanced production of multi-strange hadrons in high-multiplicity proton-proton collisions}.
\newblock {\em Nature Phys.} {\bf 2017}, {\em 13},~535--539,  \href{http://arxiv.org/abs/1606.07424}{{\normalfont [arXiv:nucl-ex/1606.07424]}}.
\newblock {\url{https://doi.org/10.1038/nphys4111}}.

\bibitem[Rath et~al.(2020)Rath, Khuntia, Sahoo, and Cleymans]{Rath:2019cpe}
Rath, R.; Khuntia, A.; Sahoo, R.; Cleymans, J.
\newblock {Event multiplicity, transverse momentum and energy dependence of charged particle production, and system thermodynamics in $pp$ collisions at the Large Hadron Collider}.
\newblock {\em J. Phys. G} {\bf 2020}, {\em 47},~055111,  \href{http://arxiv.org/abs/1908.04208}{{\normalfont [arXiv:hep-ph/1908.04208]}}.
\newblock {\url{https://doi.org/10.1088/1361-6471/ab783b}}.

\bibitem[Abelev et~al.(2013)]{ALICE:2013rdo}
Abelev, B.B.;  et~al.
\newblock {Multiplicity dependence of the average transverse momentum in pp, p-Pb, and Pb-Pb collisions at the LHC}.
\newblock {\em Phys. Lett. B} {\bf 2013}, {\em 727},~371--380,  \href{http://arxiv.org/abs/1307.1094}{{\normalfont [arXiv:nucl-ex/1307.1094]}}.
\newblock {\url{https://doi.org/10.1016/j.physletb.2013.10.054}}.

\bibitem[Tumasyan et~al.(2024)]{CMS:2023sua}
Tumasyan, A.;  et~al.
\newblock {Multiplicity and transverse momentum dependence of charge-balance functions in pPb and PbPb collisions at LHC energies}.
\newblock {\em JHEP} {\bf 2024}, {\em 08},~148,  \href{http://arxiv.org/abs/2307.11185}{{\normalfont [arXiv:nucl-ex/2307.11185]}}.
\newblock {\url{https://doi.org/10.1007/JHEP08(2024)148}}.

\bibitem[Acharya et~al.(2020)]{ALICE:2020msa}
Acharya, S.;  et~al.
\newblock {Multiplicity dependence of J/$\psi$ production at midrapidity in pp collisions at $\sqrt{s}$ = 13 TeV}.
\newblock {\em Phys. Lett. B} {\bf 2020}, {\em 810},~135758,  \href{http://arxiv.org/abs/2005.11123}{{\normalfont [arXiv:nucl-ex/2005.11123]}}.
\newblock {\url{https://doi.org/10.1016/j.physletb.2020.135758}}.

\bibitem[Ortiz et~al.(2020)Ortiz, Paz, Romo, Tripathy, Zepeda, and Bautista]{Ortiz:2020rwg}
Ortiz, A.; Paz, A.; Romo, J.D.; Tripathy, S.; Zepeda, E.A.; Bautista, I.
\newblock {Multiparton interactions in $pp$ collisions from machine learning-based regression}.
\newblock {\em Phys. Rev. D} {\bf 2020}, {\em 102},~076014,  \href{http://arxiv.org/abs/2004.03800}{{\normalfont [arXiv:hep-ph/2004.03800]}}.
\newblock {\url{https://doi.org/10.1103/PhysRevD.102.076014}}.

\bibitem[Morreale and Salazar(2021)]{Morreale:2021pnn}
Morreale, A.; Salazar, F.
\newblock {Mining for Gluon Saturation at Colliders}.
\newblock {\em Universe} {\bf 2021}, {\em 7},~312,  \href{http://arxiv.org/abs/2108.08254}{{\normalfont [arXiv:hep-ph/2108.08254]}}.
\newblock {\url{https://doi.org/10.3390/universe7080312}}.

\bibitem[Levin and Rezaeian(2010)]{Levin:2010dw}
Levin, E.; Rezaeian, A.H.
\newblock {Gluon saturation and inclusive hadron production at LHC}.
\newblock {\em Phys. Rev. D} {\bf 2010}, {\em 82},~014022,  \href{http://arxiv.org/abs/1005.0631}{{\normalfont [arXiv:hep-ph/1005.0631]}}.
\newblock {\url{https://doi.org/10.1103/PhysRevD.82.014022}}.

\bibitem[Acharya et~al.(2019)]{ALICE:2018pal}
Acharya, S.;  et~al.
\newblock {Multiplicity dependence of light-flavor hadron production in pp collisions at $\sqrt{s}$ = 7 TeV}.
\newblock {\em Phys. Rev. C} {\bf 2019}, {\em 99},~024906,  \href{http://arxiv.org/abs/1807.11321}{{\normalfont [arXiv:nucl-ex/1807.11321]}}.
\newblock {\url{https://doi.org/10.1103/PhysRevC.99.024906}}.

\bibitem[Moriggi and Machado(2022)]{Moriggi:2022xbg}
Moriggi, L.; Machado, M.V.T.
\newblock {Nuclear Modification Factor in Small System Collisions within Perturbative QCD including Thermal Effects \textdagger{}}.
\newblock {\em MDPI Physics} {\bf 2022}, {\em 4},~787--799,  \href{http://arxiv.org/abs/2207.07794}{{\normalfont [arXiv:hep-ph/2207.07794]}}.
\newblock {\url{https://doi.org/10.3390/physics4030050}}.

\bibitem[Accardi et~al.(2016)]{Accardi:2012qut}
Accardi, A.;  et~al.
\newblock {Electron Ion Collider: The Next QCD Frontier}: {Understanding the glue that binds us all}.
\newblock {\em Eur. Phys. J. A} {\bf 2016}, {\em 52},~268,  \href{http://arxiv.org/abs/1212.1701}{{\normalfont [arXiv:nucl-ex/1212.1701]}}.
\newblock {\url{https://doi.org/10.1140/epja/i2016-16268-9}}.

\bibitem[Rezaeian and Schmidt(2013)]{bCGC}
Rezaeian, A.H.; Schmidt, I.
\newblock {Impact-parameter dependent Color Glass Condensate dipole model and new combined HERA data}.
\newblock {\em Phys. Rev.} {\bf 2013}, {\em D88},~074016,  \href{http://arxiv.org/abs/1307.0825}{{\normalfont [arXiv:hep-ph/1307.0825]}}.
\newblock {\url{https://doi.org/10.1103/PhysRevD.88.074016}}.

\bibitem[Rezaeian et~al.(2013)Rezaeian, Siddikov, Van~de Klundert, and Venugopalan]{ipsat}
Rezaeian, A.H.; Siddikov, M.; Van~de Klundert, M.; Venugopalan, R.
\newblock {Analysis of combined HERA data in the Impact-Parameter dependent Saturation model}.
\newblock {\em Phys. Rev.} {\bf 2013}, {\em D87},~034002,  \href{http://arxiv.org/abs/1212.2974}{{\normalfont [arXiv:hep-ph/1212.2974]}}.
\newblock {\url{https://doi.org/10.1103/PhysRevD.87.034002}}.

\bibitem[Rezaei(2025)]{Rezaei:2024odh}
Rezaei, B.
\newblock {The nuclear shadowing effect of gluon at small x}.
\newblock {\em Nucl. Phys. A} {\bf 2025}, {\em 1053},~122971.
\newblock {\url{https://doi.org/10.1016/j.nuclphysa.2024.122971}}.

\bibitem[Krelina and Nemchik(2020)]{Krelina:2020ipn}
Krelina, M.; Nemchik, J.
\newblock {Nuclear shadowing in DIS at electron-ion colliders}.
\newblock {\em Eur. Phys. J. Plus} {\bf 2020}, {\em 135},~444,  \href{http://arxiv.org/abs/2003.04156}{{\normalfont [arXiv:hep-ph/2003.04156]}}.
\newblock {\url{https://doi.org/10.1140/epjp/s13360-020-00498-2}}.

\bibitem[Abdul~Khalek et~al.(2019)Abdul~Khalek, Ethier, and Rojo]{NNPDFnuclear}
Abdul~Khalek, R.; Ethier, J.J.; Rojo, J.
\newblock {Nuclear parton distributions from lepton-nucleus scattering and the impact of an electron-ion collider}.
\newblock {\em Eur. Phys. J. C} {\bf 2019}, {\em 79},~471,  \href{http://arxiv.org/abs/1904.00018}{{\normalfont [arXiv:hep-ph/1904.00018]}}.
\newblock {\url{https://doi.org/10.1140/epjc/s10052-019-6983-1}}.

\bibitem[Eskola et~al.(2017)Eskola, Paakkinen, Paukkunen, and Salgado]{EPPS16}
Eskola, K.J.; Paakkinen, P.; Paukkunen, H.; Salgado, C.A.
\newblock {EPPS16: Nuclear parton distributions with LHC data}.
\newblock {\em Eur. Phys. J. C} {\bf 2017}, {\em 77},~163,  \href{http://arxiv.org/abs/1612.05741}{{\normalfont [arXiv:hep-ph/1612.05741]}}.
\newblock {\url{https://doi.org/10.1140/epjc/s10052-017-4725-9}}.

\bibitem[Schienbein et~al.(2009)Schienbein, Yu, Kovarik, Keppel, Morfin, Olness, and Owens]{ncteq}
Schienbein, I.; Yu, J.; Kovarik, K.; Keppel, C.; Morfin, J.; Olness, F.; Owens, J.
\newblock {PDF Nuclear Corrections for Charged and Neutral Current Processes}.
\newblock {\em Phys. Rev. D} {\bf 2009}, {\em 80},~094004,  \href{http://arxiv.org/abs/0907.2357}{{\normalfont [arXiv:hep-ph/0907.2357]}}.
\newblock {\url{https://doi.org/10.1103/PhysRevD.80.094004}}.

\bibitem[Arneodo et~al.(1995)]{Arneodo:1995cs}
Arneodo, M.;  et~al.
\newblock {The Structure Function ratios F2(li) / F2(D) and F2(C) / F2(D) at small x}.
\newblock {\em Nucl. Phys. B} {\bf 1995}, {\em 441},~12--30,  \href{http://arxiv.org/abs/hep-ex/9504002}{{\normalfont [hep-ex/9504002]}}.
\newblock {\url{https://doi.org/10.1016/0550-3213(95)00023-2}}.

\bibitem[Adams et~al.(1995)]{Adams:1995is}
Adams, M.;  et~al.
\newblock {Shadowing in inelastic scattering of muons on carbon, calcium and lead at low x(Bj)}.
\newblock {\em Z. Phys. C} {\bf 1995}, {\em 67},~403--410,  \href{http://arxiv.org/abs/hep-ex/9505006}{{\normalfont [hep-ex/9505006]}}.
\newblock {\url{https://doi.org/10.1007/BF01624583}}.

\bibitem[Mueller(1990)]{Mueller:1989st}
Mueller, A.H.
\newblock {Small x Behavior and Parton Saturation: A QCD Model}.
\newblock {\em Nucl. Phys. B} {\bf 1990}, {\em 335},~115--137.
\newblock {\url{https://doi.org/10.1016/0550-3213(90)90173-B}}.

\bibitem[Glauber(1955)]{Glauber:1955qq}
Glauber, R.
\newblock {Cross-sections in deuterium at high-energies}.
\newblock {\em Phys. Rev.} {\bf 1955}, {\em 100},~242--248.
\newblock {\url{https://doi.org/10.1103/PhysRev.100.242}}.

\bibitem[Arneodo et~al.(1996)]{Arneodo:1996ru}
Arneodo, M.;  et~al.
\newblock {The Q**2 dependence of the structure function ratio F2 Sn / F2 C and the difference R Sn - R C in deep inelastic muon scattering}.
\newblock {\em Nucl. Phys. B} {\bf 1996}, {\em 481},~23--39.
\newblock {\url{https://doi.org/10.1016/S0550-3213(96)90119-4}}.

\bibitem[De~Jager et~al.(1974)De~Jager, De~Vries, and De~Vries]{DeJager:1974liz}
De~Jager, C.; De~Vries, H.; De~Vries, C.
\newblock {Nuclear charge and magnetization density distribution parameters from elastic electron scattering}.
\newblock {\em Atom. Data Nucl. Data Tabl.} {\bf 1974}, {\em 14},~479--508.
\newblock [Erratum: Atom.Data Nucl.Data Tabl. 16, 580--580 (1975)], {\url{https://doi.org/10.1016/S0092-640X(74)80002-1}}.

\bibitem[{De Vries} et~al.(1987){De Vries}, {De Jager}, and {De Vries}]{DEVRIES1987495}
{De Vries}, H.; {De Jager}, C.; {De Vries}, C.
\newblock Nuclear charge-density-distribution parameters from elastic electron scattering.
\newblock {\em Atomic Data and Nuclear Data Tables} {\bf 1987}, {\em 36},~495 -- 536.
\newblock {\url{https://doi.org/https://doi.org/10.1016/0092-640X(87)90013-1}}.

\bibitem[Miller et~al.(2007)Miller, Reygers, Sanders, and Steinberg]{Miller:2007ri}
Miller, M.L.; Reygers, K.; Sanders, S.J.; Steinberg, P.
\newblock {Glauber modeling in high energy nuclear collisions}.
\newblock {\em Ann. Rev. Nucl. Part. Sci.} {\bf 2007}, {\em 57},~205--243,  \href{http://arxiv.org/abs/nucl-ex/0701025}{{\normalfont [nucl-ex/0701025]}}.
\newblock {\url{https://doi.org/10.1146/annurev.nucl.57.090506.123020}}.

\bibitem[d'Enterria(2003)]{dEnterria:2003xac}
d'Enterria, D.G.
\newblock {Hard scattering cross-sections at LHC in the Glauber approach: From pp to pA and AA collisions} {\bf 2003}.
\newblock  \href{http://arxiv.org/abs/nucl-ex/0302016}{{\normalfont [nucl-ex/0302016]}}.

\bibitem[Adam et~al.(2016)]{Adam:2015kca}
Adam, J.;  et~al.
\newblock {Centrality dependence of the nuclear modification factor of charged pions, kaons, and protons in Pb-Pb collisions at $\sqrt{s_{\rm NN}}=2.76$ TeV}.
\newblock {\em Phys. Rev. C} {\bf 2016}, {\em 93},~034913,  \href{http://arxiv.org/abs/1506.07287}{{\normalfont [arXiv:nucl-ex/1506.07287]}}.
\newblock {\url{https://doi.org/10.1103/PhysRevC.93.034913}}.

\bibitem[Helenius et~al.(2013)Helenius, Eskola, and Paukkunen]{Helenius:2013bya}
Helenius, I.; Eskola, K.J.; Paukkunen, H.
\newblock {Centrality dependence of inclusive prompt photon production in d+Au, Au+Au, p+Pb, and Pb+Pb collisions}.
\newblock {\em JHEP} {\bf 2013}, {\em 05},~030,  \href{http://arxiv.org/abs/1302.5580}{{\normalfont [arXiv:hep-ph/1302.5580]}}.
\newblock {\url{https://doi.org/10.1007/JHEP05(2013)030}}.

\bibitem[Helenius et~al.(2015)Helenius, Paukkunen, and Eskola]{Helenius:2015wda}
Helenius, I.; Paukkunen, H.; Eskola, K.J.
\newblock {Nuclear PDF constraints from p+Pb collisions at the LHC}.
\newblock {\em PoS} {\bf 2015}, {\em DIS2015},~036,  \href{http://arxiv.org/abs/1509.02798}{{\normalfont [arXiv:hep-ph/1509.02798]}}.
\newblock {\url{https://doi.org/10.22323/1.247.0036}}.

\bibitem[Vitev(2007)]{Vitev:2007ve}
Vitev, I.
\newblock {Non-Abelian energy loss in cold nuclear matter}.
\newblock {\em Phys. Rev. C} {\bf 2007}, {\em 75},~064906,  \href{http://arxiv.org/abs/hep-ph/0703002}{{\normalfont [hep-ph/0703002]}}.
\newblock {\url{https://doi.org/10.1103/PhysRevC.75.064906}}.

\bibitem[Kang et~al.(2012)Kang, Vitev, and Xing]{Kang:2012kc}
Kang, Z.B.; Vitev, I.; Xing, H.
\newblock {Nuclear modification of high transverse momentum particle production in p+A collisions at RHIC and LHC}.
\newblock {\em Phys. Lett. B} {\bf 2012}, {\em 718},~482--487,  \href{http://arxiv.org/abs/1209.6030}{{\normalfont [arXiv:hep-ph/1209.6030]}}.
\newblock {\url{https://doi.org/10.1016/j.physletb.2012.10.046}}.

\bibitem[Arleo and Peign\'e(2020)]{Arleo:2020eia}
Arleo, F.; Peign\'e, S.
\newblock {Quenching of Light Hadron Spectra in $p$-A Collisions from Fully Coherent Energy Loss}.
\newblock {\em Phys. Rev. Lett.} {\bf 2020}, {\em 125},~032301,  \href{http://arxiv.org/abs/2003.01987}{{\normalfont [arXiv:hep-ph/2003.01987]}}.
\newblock {\url{https://doi.org/10.1103/PhysRevLett.125.032301}}.

\bibitem[Arleo et~al.(2020)Arleo, Cougoulic, and Peign\'e]{Arleo:2020hat}
Arleo, F.; Cougoulic, F.; Peign\'e, S.
\newblock {Fully coherent energy loss effects on light hadron production in pA collisions}.
\newblock {\em JHEP} {\bf 2020}, {\em 09},~190,  \href{http://arxiv.org/abs/2003.06337}{{\normalfont [arXiv:hep-ph/2003.06337]}}.
\newblock {\url{https://doi.org/10.1007/JHEP09(2020)190}}.

\end{thebibliography}

\end{document}